\documentclass[fleqn,10pt]{wlpeerj}

\usepackage{graphicx}
\usepackage{hyperref}
\usepackage{amsmath}
\usepackage{amsthm}
\newtheorem{lemma}{Lemma}

\usepackage{caption}
\usepackage{subcaption}

\usepackage[linesnumbered,ruled,vlined]{algorithm2e}

\SetCommentSty{mycommfont}

\usepackage{courier}

\usepackage{listings}
\definecolor{cppColorBackground}{rgb}{1.,1.,1.}
\definecolor{cppColorComment}{rgb}{0.0,0.47,.8}
\definecolor{cppColorLine}{rgb}{0.6,0.6,0.6}
\definecolor{cppColorString}{rgb}{0,0.501,145}
\definecolor{cppColorKey}{rgb}{0.8,0.5,0}
\definecolor{cppColorDigit}{rgb}{0,0,0.5}
\title{Exploiting ray tracing technology through OptiX to compute particle interactions with cutoff in a 3D environment on GPU}

\author[1,2]{David Algis}
\author[3]{B\'erenger Bramas}
\affil[1]{University of Poitiers, XLIM, France}
\affil[2]{Studio Nyx}
\affil[3]{Inria Nancy, ICube Laboratory, University of Strasbourg, France}
\corrauthor[.]{B\'erenger Bramas}{Berenger.Bramas@inria.fr}

\begin{abstract}
Computing on graphics processing units (GPUs) has become standard in scientific computing, allowing for incredible performance gains over classical CPUs for many computational methods.
As GPUs were originally designed for 3D rendering, they still have several features for that purpose that are not used in scientific computing.
Among them, ray tracing is a powerful technology used to render 3D scenes.
In this paper, we propose exploiting ray tracing technology to compute particle interactions with a cutoff distance in a 3D environment.
We describe algorithmic tricks and geometric patterns to find the interaction lists for each particle.
This approach allows us to compute interactions with quasi-linear complexity in the number of particles without building a grid of cells or an explicit kd-tree.
We compare the performance of our approach with a classical approach based on a grid of cells and show that, currently, ours is slower in most cases but could pave the way for future methods.
\end{abstract}

\begin{document}

\flushbottom
\maketitle
\thispagestyle{empty}

\section{Introduction}

High-performance computing (HPC) is a key technology in scientific computing. 
Since the early 2000s, the use of graphics processing units (GPUs) has become standard in HPC, and they now equip many of the fastest supercomputers~\footnote{\url{https://top500.org/}}. 
GPUs are massively parallel processors that allow computations to be performed on thousands of cores, making them perfectly suited for inherently parallel problems. 
The use of GPUs in scientific computing has led to incredible performance gains in many fields, such as molecular dynamics, fluid dynamics, astrophysics, and machine learning.

Most scientific computing applications that use GPUs are based on the CUDA~\footnote{\url{https://developer.nvidia.com/cuda-toolkit}} programming model, and to a lesser extent, the OpenCL programming model. 
They do not exploit all the features of GPUs, particularly those dedicated to 3D rendering. Among these features, ray tracing is a powerful technology used to render 3D scenes. 
In this method, rays are cast from the camera to the scene, and the intersections of the rays with the objects in the scene are computed to generate the colors of the pixels in the image. 
For example, the NVIDIA GeForce RTX 4090, as a single consumer-grade GPU, demonstrates the raw power of this technology by rendering complex 3D scenes at 87 frames per second in 4K resolution (3840 x 2160 pixels), handling millions of triangles per frame~\cite{nvidia2024ada}.

In this paper, we are interested in evaluating how ray tracing technology could be used to compute particle interactions in a 3D environment. 
Our motivation is twofold: we want to evaluate if it is possible to use ray tracing technology, and we want to create the algorithmic patterns needed for that purpose.

With these aims, we focus on the computation of particle interactions in a 3D environment, which is a common problem in scientific computing. 
When the potential of the interaction kernel decreases exponentially with distance, the interactions can be computed with a cutoff distance, i.e., the interactions are only computed between particles that are closer than a given distance, achieving less but still satisfactory accuracy. 
This allows the complexity of the interactions to drop from $O(N^2)$ to $O(N)$, where $N$ is the number of particles if the cutoff distance is small enough. 
Classical methods to compute such interactions are based on the use of a grid of cells or an explicit kd-tree to quickly find the interaction lists for each particle. 
In this paper, we aim to avoid using such data structures and instead exploit ray tracing technology.

The contributions of this paper are as follows:
\begin{itemize}
\item We propose a method to compute particle interactions with a cutoff distance in a 3D environment based on ray tracing technology.
\item We describe two algorithmic techniques based on geometric patterns to find the interaction lists for each particle using real intersections.
\item We compare the performance of our approach with a classical approach based on a grid of cells and with an existing method that relies on ray tracing.
\end{itemize}

The rest of the paper is organized as follows.
In Section~\ref{sec:prerequisites}, we present the prerequisites.
In Section~\ref{sec:related_work}, we review the related work.
In Section~\ref{sec:proposed_approach}, we present our proposed approach.
In Section~\ref{sec:performance_study}, we present the performance study.
Finally, we conclude in Section~\ref{sec:conclusion}.

%%%%%%%%%%%%%%%%%%%%%%%%%%%%%%%%%%%%%%%%%%%%%%%%%%%%%%%%%%%%%%%%%%%%%%%%%%%%%%%%%
%%%%%%%%%%%%%%%%%%%%%%%%%%%%%%%%%%%%%%%%%%%%%%%%%%%%%%%%%%%%%%%%%%%%%%%%%%%%%%%%%

\section{Prerequisites}
\label{sec:prerequisites}

\subsection{Particle Interactions}
\label{subsec:particles_interactions}
Computing the interactions between $N$ particles in a 3D environment is a common problem in scientific computing. For example, this is essential in fluid simulations using smoothed particle hydrodynamics~\cite{koschier2020smoothed} and in molecular dynamics for simulating forces between atoms~\cite{badar2022molecular}.
These interactions are usually modeled by a potential function that depends on the distance between the particles.
A straightforward way to compute the interactions is to evaluate the potential function for all pairs of particles.
However, the potential function can be short-range or long-range.
When the potential function is short-range, the interactions can be computed with a cutoff distance, i.e., the interactions are only computed between particles that are closer than a given distance.
This reduces the complexity of the interactions from $O(N^2)$ to $O(N)$, where $N$ is the number of particles, but this is only possible if we have an efficient way to find the particles' neighbors.
Moreover, the positions of the particles are usually updated after each computation step.
Consequently, the system used to find the interactions between the particles should be updated at each iteration of the simulation.

A possible solution to get the interaction list is to build a grid of cells mapped over the simulation space, where each cell contains the particles that are inside.
The cells have a width equal to the cutoff distance $C$.
Then, for each particle, the interactions are computed with the particles that are in the same cell and in the neighboring cells.
Building the grid of cells and finding the interaction lists for each particle can be done in $O(N)$: we start by computing the cell index for each particle, then we order the particles in a new array by assigning a unique index per particle using atomic operations, and finally, we move the particles to a new array~\cite{algis:hal-04621128}.
Each of these three operations can be implemented with a parallel loop over the $N$ particles.

\subsection{Graphics Processing Units}

Graphics processing units (GPUs) are massively parallel processors that allow computations to be performed on thousands of cores. 
The hardware design of GPUs has been optimized for graphics rendering, particularly for the rendering of 3D scenes. 
To this end, GPUs have features dedicated to 3D rendering, such as texture mapping, rasterization, and ray tracing. 
Internally, GPUs are organized in a hierarchy of processing units, including streaming multiprocessors (SMs), warp schedulers, and execution units.

NVIDIA has proposed the CUDA programming model to develop parallel applications for GPUs. 
CUDA is designed to express algorithms in a way that can be mapped to the GPUs' hardware organization. 
For instance, thread blocks are distributed across SMs, and threads are executed in warps. 
There are also keywords and functions to use shared memory, constant memory, and texture memory. 
Thus, creating optimized applications for GPUs requires an understanding of the hardware architecture of GPUs and potentially adapting algorithms to their specificities.

\subsection{Ray Tracing}

Among the many features of GPUs dedicated to 3D rendering, ray tracing stands out as a powerful technology widely used in video games for rendering realistic 3D scenes ~\cite{ign2023Top5Games}. It is a hardware-accelerated technique that computes the interactions of rays with objects in a scene. Ray tracing generates pixel colors in an image by casting rays from the camera into the scene. For each intersection of a ray with an object, the pixel's color is determined based on the object's material properties and lighting conditions. Rays can also be reflected or refracted by objects, or continue through non-opaque surfaces, enabling recursive computation of interactions for enhanced realism.

Typically, ray tracing kernels are implemented within the shaders of the graphics pipeline. Shaders are small programs executed on the GPU for each pixel in an image, often written in specialized languages such as GLSL (for OpenGL) or HLSL (for DirectX). These shaders run in parallel on the GPU, allowing for concurrent computation of ray-object interactions. NVIDIA introduced a method to utilize ray tracing within the CUDA programming model through its OptiX library~\cite{10.1145/1778765.1778803}. With OptiX, developers retain the flexibility of CUDA programming while leveraging ray tracing technology, albeit with some constraints on kernel implementation.

\begin{figure}[htb!]
  \centering
    \includegraphics[width=\textwidth, height=.18\textheight, page=1, keepaspectratio]{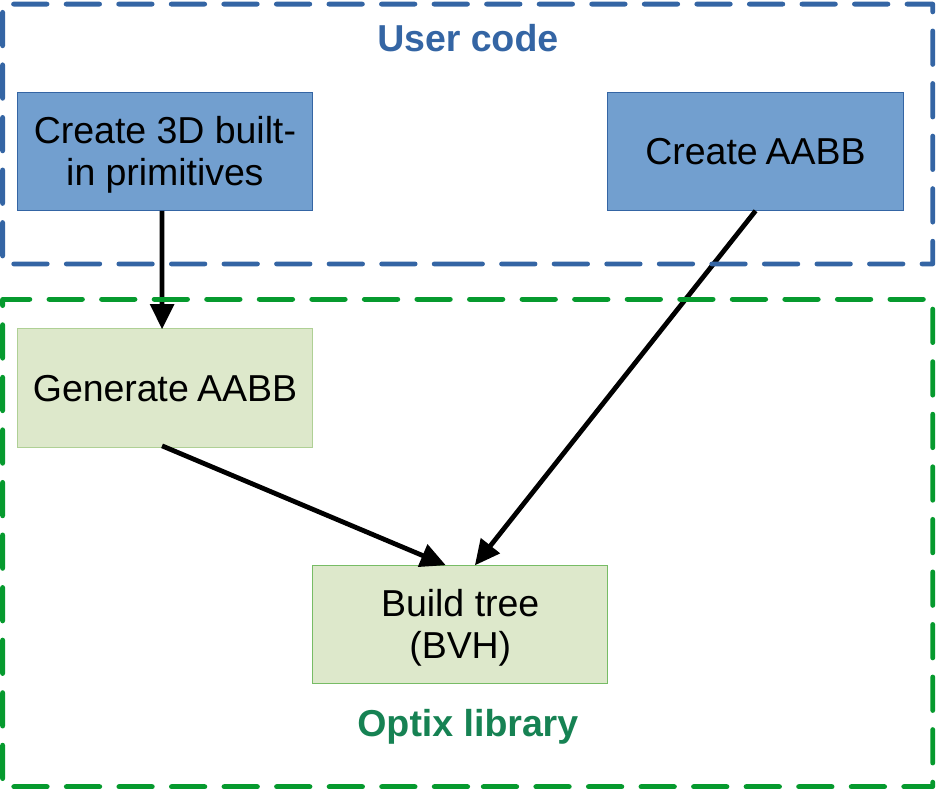}
\caption{Schematic view of BVH tree creation. The OptiX library requires a list of encompassing axis-aligned bounding boxes (AABBs), which can either be directly provided or generated from a list of basic primitives.}
\label{fig:aabbtohbv}
\end{figure}

In ray tracing, scenes are represented using geometric primitives such as triangles or spheres, which are encapsulated within bounding volumes to enhance computational efficiency~\cite{Shirley2019Ray}. A common bounding volume is the Axis-Aligned Bounding Box (AABB), a rectangular box aligned with the coordinate axes and defined by its minimum and maximum bounds $(x_{min},y_{min},z_{min})$ and $(x_{max},y_{max},z_{max})$.
AABB-ray intersection tests are conducted by calculating $t_{min}$ and $t_{max}$ for each axis and ensuring that overlaps exist across all axes, efficiently determining whether the ray intersects the box.

To further optimize performance, AABBs are organized into a Bounding Volume Hierarchy (BVH), a tree-like data structure where each node represents an AABB. Internal nodes group child AABBs, while leaf nodes encapsulate the actual primitives. In OptiX, users can either provide a list of basic primitives or directly supply a list of AABBs, as illustrated in Figure~\ref{fig:aabbtohbv}.

During ray traversal, the algorithm performs intersection tests at each BVH node. If a ray misses an AABB, the entire subtree beneath that node is skipped, reducing unnecessary computations. When a ray reaches a leaf node, precise intersection tests are conducted with the enclosed primitives, and the closest intersection is recorded. This process reduces the complexity of ray-scene intersections from linear to logarithmic by pruning large portions of the scene, ensuring only relevant branches of the BVH are explored.

\begin{figure}[htb!]
  \centering
    \includegraphics[width=\textwidth, height=.3\textheight, page=1, keepaspectratio]{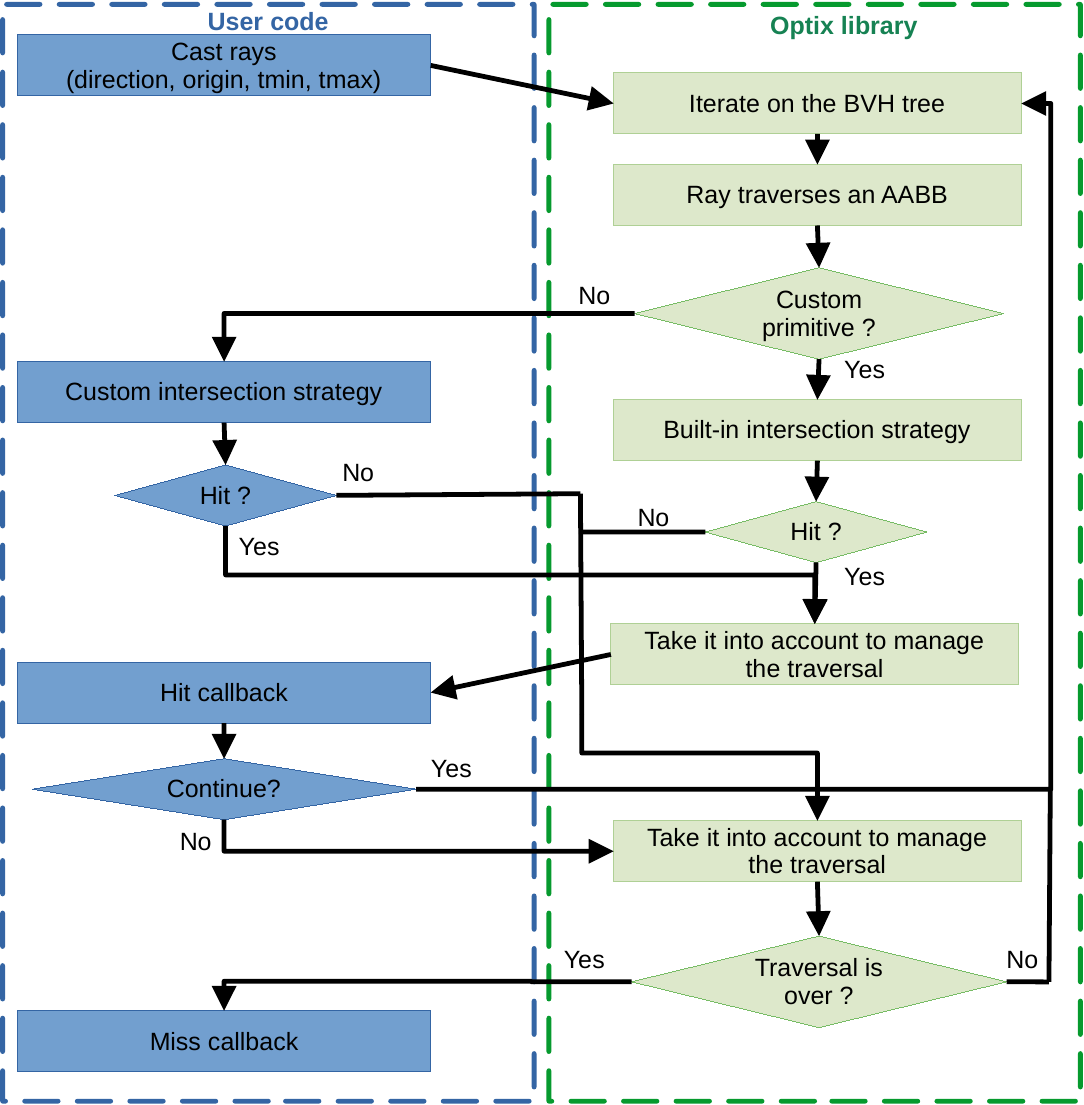}
\caption{Schematic view of the OptiX ray tracing workflow.
The user casts rays in the desired direction, and OptiX manages the BVH tree traversal and potential interactions.
When a ray enters an enclosing bounding box (AABB), a callback function is invoked to determine whether the ray actually intersects the corresponding primitive, based on a built-in or user-defined Intersection Shader (IS), referred to as the intersection strategy.
For custom primitives, the user must implement the IS.
If the ray does not intersect or is marked to continue, the process is repeated for other primitives.
The traversal stops when a final intersection is found or no more primitives can be intersected, triggering a call to the miss callback.}
\label{fig:optixrt}
\end{figure}

The final output of the ray tracing algorithm, typically the closest hit point, is used for shading or further processing to determine the visual appearance of the scene. This hierarchical approach, combined with the use of AABBs and BVH, ensures that ray tracing remains computationally feasible even for complex scenes containing millions of primitives. Figure~\ref{fig:optixrt} provides a summary of the different operations in ray tracing using OptiX.

%%%%%%%%%%%%%%%%%%%%%%%%%%%%%%%%%%%%%%%%%%%%%%%%%%%%%%%%%%%%%%%%%%%%%%%%%%%%%%%%%
%%%%%%%%%%%%%%%%%%%%%%%%%%%%%%%%%%%%%%%%%%%%%%%%%%%%%%%%%%%%%%%%%%%%%%%%%%%%%%%%%

\section{Related Work}
\label{sec:related_work}

\subsection{Neighbor Search on GPU}

The main work on physical simulation of particle interactions on GPU has been proposed by ~\cite{nylons2007fast}, to compute the gravitational potential.
They described an efficient implementation using shared memory that became a standard implementation on GPU.
As mentioned in Section \ref{subsec:particles_interactions}, grid-based and kd-tree structures are widely utilized for neighbor search operations on GPUs. For grid-based approaches, a detailed description can be found in our previous study \cite{algis:hal-04621128}, which focuses on scenarios with few particles per cell and demonstrates that utilizing shared memory often does not yield significant benefits. Regarding kd-tree implementations, \cite{garcia2008fast} present a method for fast k-nearest neighbor searches using GPUs, highlighting the efficiency of kd-trees in high-dimensional spaces.

\subsection{Ray Tracing in Computer Graphics}

Ray tracing has been widely applied in particle-based representations, primarily in the domain of computer graphics and rendering. \cite{10.1145/3687934} propose a method for efficiently ray tracing Gaussian particles to enable advanced rendering effects such as shadows, reflections, and depth of field in dense particle scenes, with applications in novel-view synthesis and visual realism. Similarly, \cite{lindau2020hardware} explores hardware-accelerated ray tracing for rendering particles that cast shadows, focusing on evaluating the performance of a prototype system. While these works demonstrate the utility of ray tracing in particle-based scenes, their focus lies in rendering and visualization, contrasting with our application of ray tracing for neighbor search in physical simulations.

\subsection{OptiX in Scientific Computing}

The NVIDIA OptiX framework has shown potential for diverse applications in scientific computing. Blyth et al. utilized OptiX to enable high-performance optical photon simulations in particle physics. This approach reduced memory and computation overheads by using GPU-based culling of photon hits, handling millions of photons in complex geometries~\cite{blyth2019opticks,blyth2021integration}. OptiX has been utilized as a flexible and high-performing tool for optical 3D modeling, enabling virtual measurements of sample surfaces by tracing over 1 billion light rays per image and comparing simulated results with those from physical measuring devices~\cite{keksel2023scientific}.

While these applications highlight OptiX's potential in handling computationally intensive tasks, they primarily focus on modeling physical processes rather than using ray tracing for spatial queries or particle interactions. Our work bridges this gap by leveraging OptiX for neighbor detection and particle interaction calculations, building on its demonstrated strengths in scientific modeling to expand its applicability into the domain of particle-based simulations.

\subsection{OptiX for Neighbor Search}

Using OptiX to find neighbors between elements has already been explored in several works.

%I. Evangelou et al.~
\cite{Evangelou2021RadiusSearch} introduced a novel approach to spatial queries, particularly radius search, by leveraging GPU-accelerated ray-tracing frameworks with OptiX. Instead of traditional spatial data structures like kd-trees, the authors proposed mapping the radius-search task to the ray-tracing paradigm by treating query points as primitives within a bounding volume hierarchy (BVH). In the proposed method, a ray used in the radius-search operation is essentially infinitesimal in extent, and its purpose is not to compute a traditional intersection but to test whether the origin of the ray (the query point) is within the bounding volume of a "sphere" surrounding each sample point. For that purpose, the authors reimplemented the function that tests if a ray intersects a sphere to control how the tests are performed. This approach enabled significant performance gains in dynamically updated datasets.

%Yuhao Zhu~
\cite{10.1145/3503221.3508409} further advanced the application of ray tracing for neighbor search by introducing optimizations for mapping the neighbor search problem onto the ray-tracing hardware available in modern GPUs. The author identified two key performance bottlenecks: unmanaged query-to-ray mapping, which led to control-flow divergences, and excessive tree traversals stemming from monolithic BVH construction. To address these issues, they proposed query scheduling and partitioning strategies that exploit spatial coherence and reduce BVH traversal time. Their experiments demonstrated substantial speedups ranging from $2.2$ to $65.0$ over existing GPU neighbor search libraries. Their work highlights the potential of using ray-tracing hardware not only for rendering but also for efficient spatial queries.

% Shiwei Zhao et al.~
\cite{raygpu} extended the use of ray tracing for particle-based simulations by converting neighbor search into a ray tracing problem. Each particle was represented as a bounding box, similar to previous works, with tiny rays emitted to detect intersections and identify neighboring particles. By leveraging NVIDIA's RT cores alongside CUDA cores, they demonstrated $10\%$ to $60\%$ performance improvements over traditional cell-based methods in various particle-based simulations, including discrete element methods and smooth particle hydrodynamics. Their approach underscores the versatility of ray-tracing cores for accelerating computationally intensive neighbor search tasks across different domains.

These works collectively demonstrate the potential of using OptiX and ray-tracing hardware for efficient spatial queries, providing a foundation for further exploration in GPU-accelerated computational methods. However, these approaches lack a direct integration of physical interaction computations within the ray-tracing framework. Our approach addresses this gap by embedding cutoff distances directly into the intersection tests. Additionally, we introduce two novel methods for neighbor detection that utilize actual intersection computations, accommodating scenarios where custom IS are unsupported.

%%%%%%%%%%%%%%%%%%%%%%%%%%%%%%%%%%%%%%%%%%%%%%%%%%%%%%%%%%%%%%%%%%%%%%%%%%%%%%%%%
%%%%%%%%%%%%%%%%%%%%%%%%%%%%%%%%%%%%%%%%%%%%%%%%%%%%%%%%%%%%%%%%%%%%%%%%%%%%%%%%%

\section{Proposed Solutions}

\subsection{Overview}
\label{sec:proposed_approach}

The core idea of our approach is to represent particles using geometric primitives and to use ray tracing to find the neighbors of each particle by detecting intersections with these primitives. 
With this aim, we investigate three possibilities:

\begin{itemize}
  \item Custom AABB: in this case, we directly provide the list of englobing bounding boxes.
        Therefore, we also have to provide our own IS in which we do not compute real intersections but only check if a ray is inside an AABB.
        This strategy is the closest one to the state of the art.
  \item Spheres: in this case, we use built-in sphere primitives, which allows us to use the built-in IS that check if a ray really intersect with the surface of a sphere.
        However, we had to create our own geometric algorithm to make it work.
  \item Triangles: in this case, we use built-in triangle primitives to create squares, which allows to use the built-in IS that check if a ray really intersect with the surface of a triangles.
        As for the spheres, we had to create our own geomatric algorithm to make it work.
\end{itemize}

The spherical approach is simpler and more intuitive than using squares,
but we are interested in evaluating if the squares made of triangles is more efficient as it is the most used primitive in 3D rendering.
We also want to evaluate if the built-in IS is more efficient than our custom IS.

In the cases that rely on built-in IS, we use the following algorithmic pattern:
\begin{enumerate}
\item We build a geometric representation for each particle.
\item We cast rays from each particle in specific directions to find the neighbors, depending on the representation.
\item We filter the intersections to avoid computing the same interaction multiple times.
\item We compute the interactions between the particles that are closer than $C$.
\end{enumerate}
To reduce overhead, we aim to use as few rays as possible and ensure they do not intersect with too many particles that are not within the distance $C$.

In terms of implementation, we use the OptiX library to develop the ray tracing kernels, which can be used in conjunction with the CUDA programming model. 
Specifically, in the OptiX API, we create a scene by providing a list of geometric primitives. 
We then provide a CUDA kernel that launches the rays, where each ray has an origin, a direction, a starting point, and an endpoint. 
Usually, one CUDA thread is used to launch one ray. 
Finally, a callback is invoked by OptiX when a ray intersects with a primitive or if no intersection is found between the starting and ending points (in 3D rendering, this usually means that the background color should be used).

The data accessible from the callback is limited. 
OptiX built-in IS can provide information about the intersection, such as the intersection point, the normal on the surface, the distance from the ray's origin, and the index of the primitive that was hit. 
Additionally, the user can pass information from the CUDA kernel that launches the rays to the callback using a payload. 
A payload is a user-defined data structure that is passed along with a ray as it traverses the scene. 
It allows the ray to carry information that can be read or modified. 
The number of payload variables is limited (usually 16 32-bit integers in recent versions).

When the hit callback is invoked, and we want the ray to continue, there are two possibilities. 
The first is to launch a new ray from the intersection point in a new direction. 
This is done by storing the intersection distance in a payload variable, returning from the callback, and then launching a new ray from the intersection point using a loop in the CUDA kernel to reach the desired distance. 
The second possibility is to inform OptiX that we want to continue the traversal by calling a corresponding function in the intersection callback. This second approach is generally more efficient, as it avoids the overhead associated with launching new rays and and should be favored in practice.

Additionally, we cannot allocate memory in the callbacks, so we cannot build complex data structures, such as lists, to store intersection lists. 
Consequently, if we want to filter the intersections, we cannot fill an array with indices and check if an index exists in the array to ensure uniqueness; instead, we must use geometric properties to filter the intersections.

In the remainder of the section, we consider that the target particle is the particle for which we want to find the neighbors.

%%%%%%%%%%%%%%%%%%%%%%%%%%%%%%%%%%%%%%%%%%%%%%%%%%%%

\subsection{Spherical Representation}
\label{sec:spherical_representation}

In this section, we consider the case where the particles are represented by spheres and we use the built-in IS. 
In OptiX, a sphere is a geometric primitive defined by its center and radius, and multiple spheres of the same radius can be instantiated in the scene, which is the approach we use.
We have the following objectives:
\begin{enumerate} 
\item Expressing the radius of the spheres depending on the cutoff distance;
\item Defining the origins and directions of the rays;
\item Providing a mechanism to filter the intesection when the rays intersect multiple times with the same sphere.
\end{enumerate}

In our model, we will use three rays for each particle, one in each direction of the coordinate system.
Consider a sphere of radius $C$ centered at the origin in a three-dimensional space.
The points on this sphere that are at the farthest distance from the three coordinate axes are located in the corners of a cube inscribed within the sphere. 
For instance, one such point at distance $C$ from the origin lies in the direction $(1, 1, 1)$. 
These 8 points, corresponding to the vertices of the cube, are all at a distance $C$ from the origin, and we want to know how far they are from the coordinate axes. 
This can be calculated as follows: since the points have coordinates where $|x| = |y| = |z|$, we use the equation of the sphere $x^2 + y^2 + z^2 = C^2$. 
If we take the point for which $x = y = z$, it gives $3x^2 = C^2$, resulting in $x = \frac{C}{\sqrt{3}}$. 
Therefore, each of these points is at a distance of $\frac{C \times \sqrt{2} }{\sqrt{3}}$ from any of the three coordinate axes.

\begin{figure}[htb!]
    \centering
    \begin{subfigure}[b]{0.49\linewidth}
        \centering
        \includegraphics[width=\textwidth, height=.2\textheight, page=1, keepaspectratio]{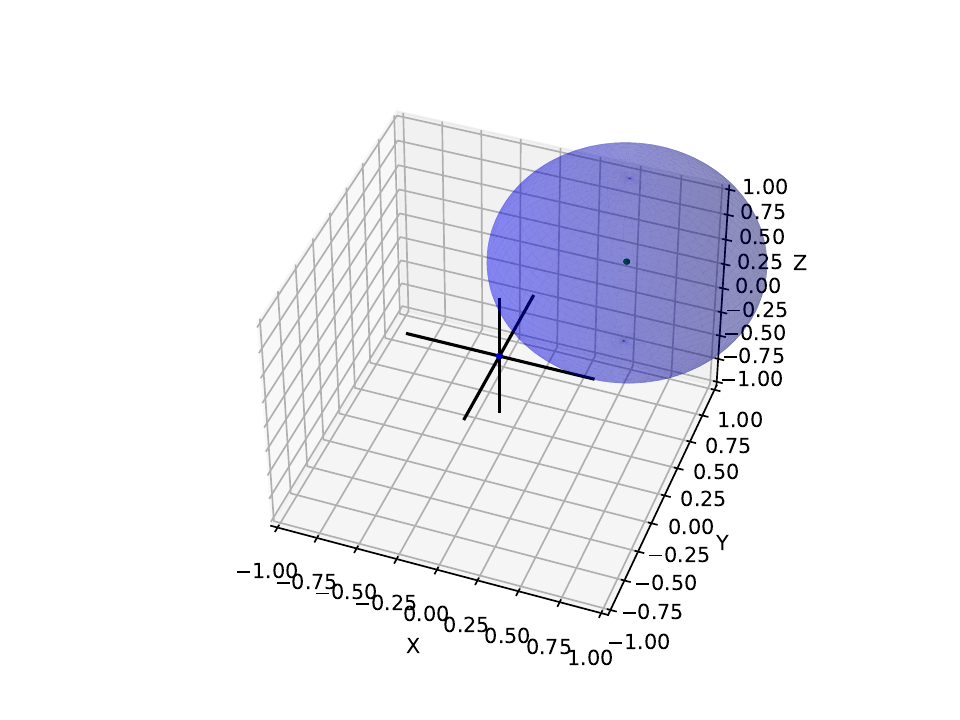}
  \end{subfigure}
  \hfill
  \begin{subfigure}[b]{0.49\linewidth}
      \centering
      \includegraphics[width=\textwidth, height=.2\textheight, page=1, keepaspectratio]{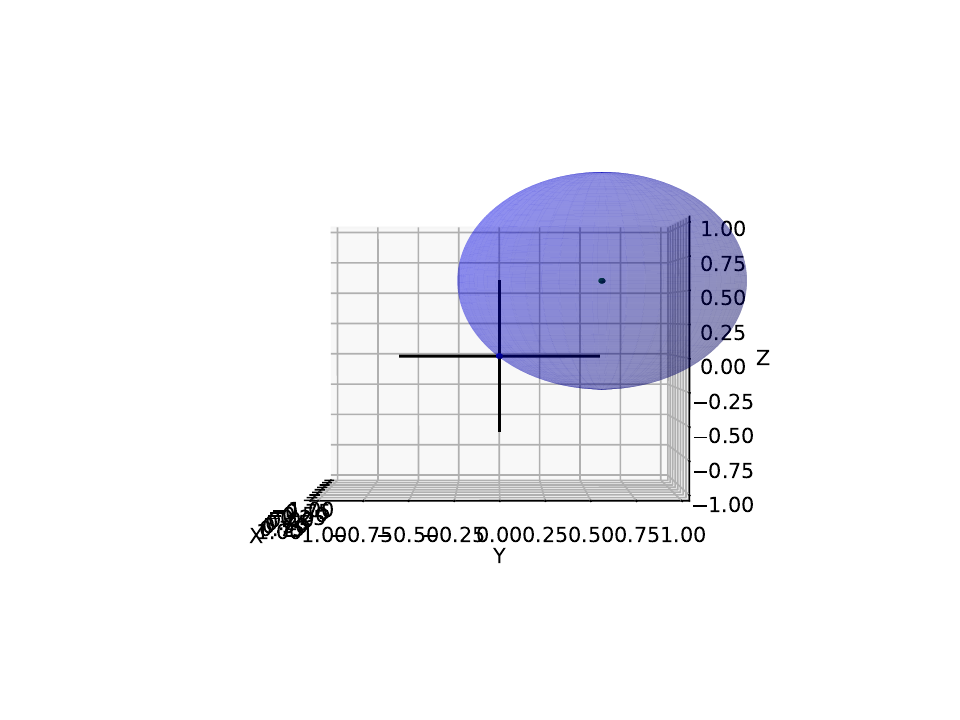}
  \end{subfigure}
  \caption{3D Spherical representation of a source and target particles distance from $C=1$.
           The source sphere has a radius of $\frac{\sqrt{2}}{\sqrt{3}}$ and the rays, represented by segments, are of length $\frac{1}{\sqrt{3}}$ in each direction.
           In this case, the sphere could englobe the rays.}
  \label{fig:sphere}
  \end{figure}

We use this information to define the radius of the spheres and the length of our rays.
The radius is set to $r = \frac{C \times \sqrt{2} }{\sqrt{3}}$.
In this scenario, it is sufficient that the rays go up to $l = \frac{C}{\sqrt{3}}$ in each direction relatively to the particle's position (so a single ray goes from $-l$ to $l$).
We provide a simplified 3D rendering of the spheres in Figure~\ref{fig:sphere} that illustrate our model.

However, if $l < r$, there are positions where the sphere could simply englobe the rays, and we would miss some intersections (when the source and target are closer than $r-l$ in the three dimensions).
Therefore, we set $l = r$ and add an $\epsilon$ to the radius of the sphere to ensure that the rays intersect with the sphere in all cases, 
obtaining $r = \frac{C \times \sqrt{2} }{\sqrt{3}} + \epsilon$ and $l = \frac{C \times \sqrt{2} }{\sqrt{3}}$.
The $\epsilon$ is a small value such that it must be impossible that two particles are closer than $\epsilon$, or some intersections will be missed (see Appendix~\ref{app:proof} for more details).

\begin{figure}[htb!]
\centering
\includegraphics[width=\textwidth, page=1, keepaspectratio, clip, trim=0cm 6cm 6cm 0cm]{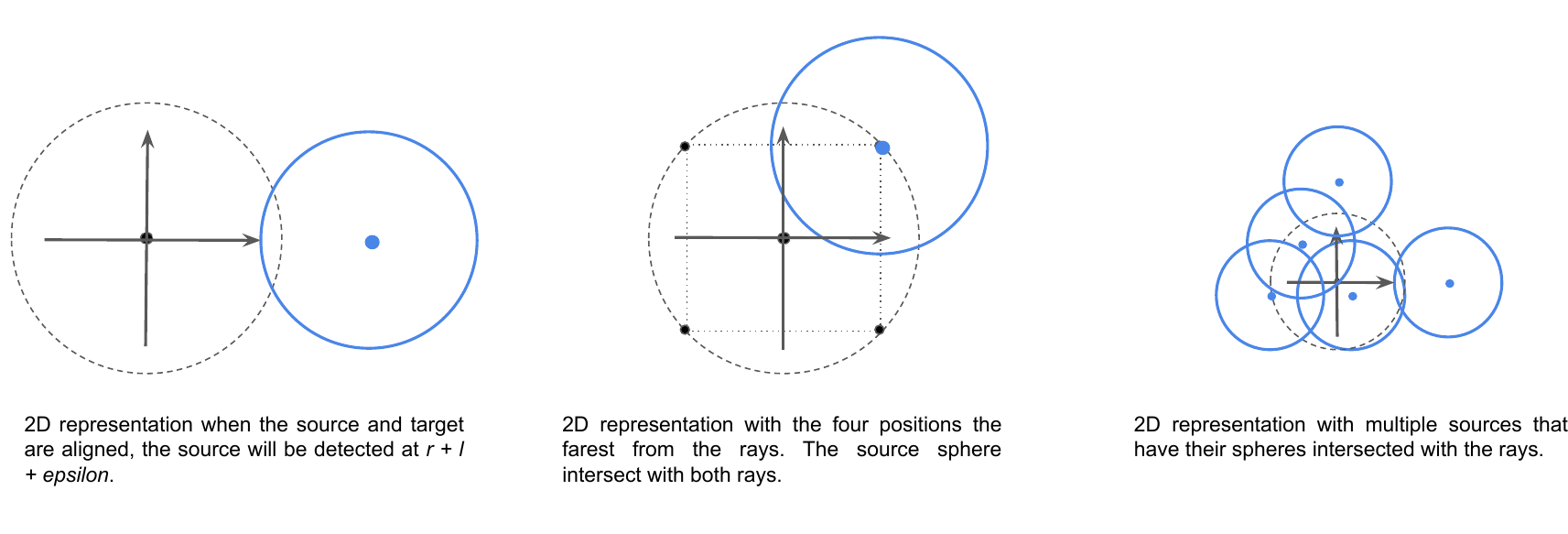}
\caption{2D Spherical representation of the particles in three different cases.
         In the first one (left), we what will be the positioned of the farthest source particle and how it will be detected by the ray.
         In the second one (center), we show the case where the source particle is the farthest from the rays (it also shows that in 2D the sphere radius could be smaller).
         In the last one (right), we show different source particles with their spheres and the rays that will intersect with them.}
\label{fig:spherical_representation}
\end{figure}

When the source and target particles are perfectly aligned on one axis, the ray will intersect for particles distant from the extremity by $r + \epsilon$.
Therefore, two particle distant from $l + r + \epsilon$ can interact.
This case and any intermediate situation where the source/target are actually too far can easily be filtered by checking the distance.
In Figure~\ref{fig:spherical_representation}, we provide a 2D representation of the particles using spheres on intricate cases.

However, each ray can potentially intersect with a sphere several times, and the same sphere can be intersected by several rays of the same target particle, so we need to filter them. 
It is impossible to maintain a global list that all the rays can access to check if an intersection has already been found, or even a single list per ray to ensure that it does not intersect with the same sphere. 
Therefore, we must do this based on geometric rules as described in Algorithm~\ref{alg:sphere_intersection}.
When we detect an intersection, we get the position of the source and target particles and first check they are within the cutoff distance. 
Then, we check if the closest ray to the source particle is the current one, and if yes, we can proceed with the computation (see Appendix~\ref{app:proof} for more details).

\begin{algorithm}[htb!]
  \DontPrintSemicolon
  \SetKwFunction{optixGetSphereData}{optixGetSphereData}
  \SetKwFunction{getPayloadPartPos}{getPayloadPartPos}
  \SetKwFunction{getPayloadC}{getPayloadC}
  \SetKwFunction{abs}{abs}
  \SetKwFunction{getClosestAxis}{getClosestAxis}
  \SetKwFunction{optixGetWorldRayDirection}{optixGetWorldRayDirection}
  
  \KwData{Optix variables}
  \KwResult{Callback when a ray intersect with a sphere}
  \textbf{Function} callback()\;
  \Begin{
    \tcc{Get the center of the sphere (source position)}
    $q \gets$ \optixGetSphereData()\;
    
    \tcc{Current particle position (target position)}
    $point \gets \getPayloadPartPos()$\;
    \tcc{Compute differences and distances}
    $diff\_pos \gets \{\abs(point.x - q.x), \abs(point.y - q.y), \abs(point.z - q.z)\}$\;
    $diff\_pos\_squared \gets \{diff\_pos.x^2, diff\_pos.y^2, diff\_pos.z^2\}$\;
    $dist\_squared \gets diff\_pos\_squared.x + diff\_pos\_squared.y + diff\_pos\_squared.z$\;
    \tcc{Get the cutoff distance from payload}
    $c \gets \getPayloadC()$\;

    \tcc{Ensure it is in the cutoff distance}
    \If{$dist\_squared < c^2$}{
        $dist\_axis\_squared \gets \{diff\_pos\_squared.y + diff\_pos\_squared.z, diff\_pos\_squared.x + diff\_pos\_squared.z, diff\_pos\_squared.x + diff\_pos\_squared.y\}$\;
        $ray\_dir \gets \optixGetWorldRayDirection()$\;
        
        $closest\_axis \gets \getClosestAxis(dist\_axis\_squared)$\;
        $closest\_axis\_is\_ray\_dir \gets (ray\_dir == closest\_axis)$\;
        
        \tcc{Ensure this ray and this intersection are the good one}
        \If{$closest\_axis\_is\_ray\_dir$}{
          \tcc{Call computation kernel}
        }
    }
  }
  \caption{Sphere intersection callback}
  \label{alg:sphere_intersection}
\end{algorithm}

%%%%%%%%%%%%%%%%%%%%%%%%%%%%%%%%%%%%%%%%%%%%%%%%%%%%

\subsection{Double Squares Representation}
\label{sec:triangular_representation}

Most 3D rendering applications use triangles to represent the objects in the scene, which motivated us to create a second model that relies on triangles instead of spheres. 
Of course, building the AABB representations and the HBV tree is not expected to be faster than the sphere, but the hardware modules and built-in could be more optimize for the triangles, as it is more common primitives.
In our model, we use four triangles to draw two squares, which are positioned opposite each other. 
Each square has a width of $C + \epsilon$ and is positioned at a distance of $C/2$ along the X-axis relative to the particle, one in each direction. 
We provide Code~\ref{lst:parttopos} in Appendix~\ref{app:postotriangles}, which shows how we generate the triangles from the particles' positions.

\begin{figure}[htb!]
\centering
\includegraphics[width=\textwidth, page=2, keepaspectratio, clip, trim=0cm 5cm 6cm 0cm]{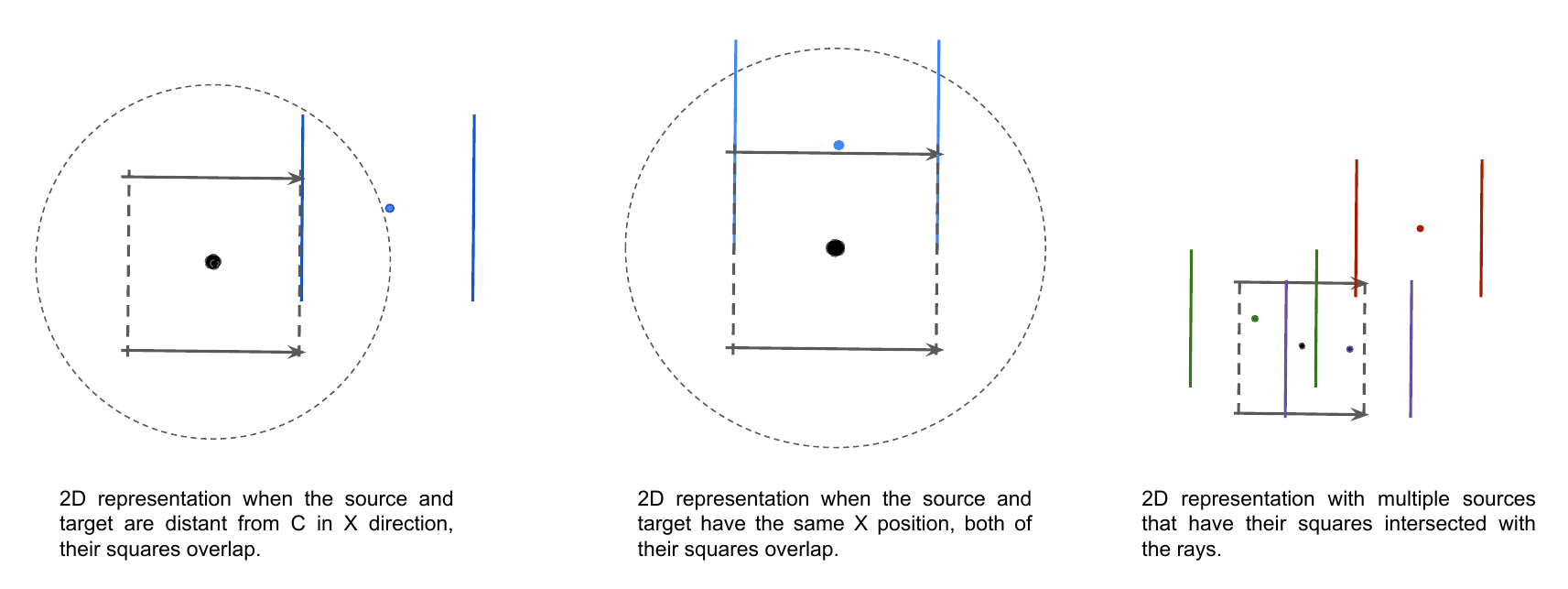}
\caption{Double squares (line) representation of the particles.
         On the left, we show how the squares can overlap for particles that are too far, but which can be easily filtered with the distance.
         In the middle, we show how particles that have the same $x$ coordinate can have their squares that overlap.
         On the right, we show different source particles with their squares and the rays that will intersect with them.}
\label{fig:triangular_representation}
\end{figure}

We then launch four rays, all with the same length and direction, but positioned at the corners of the squares.
Each ray is positioned at $-C/2$ from the center of the square and has a length of $C + \epsilon$.
The $\epsilon$ is used to ensure that when the source and target particles have the same $x$ coordinate and their squares overlap, the rays cross the triangles.
We provide in Figure~\ref{fig:triangular_representation} the 2D representation of the particles using squares (which are lines in 2D).

\begin{figure}[htb!]
  \centering
  \begin{subfigure}[b]{0.3\linewidth}
    \centering
    \includegraphics[width=\textwidth, height=.2\textheight, page=1, keepaspectratio]{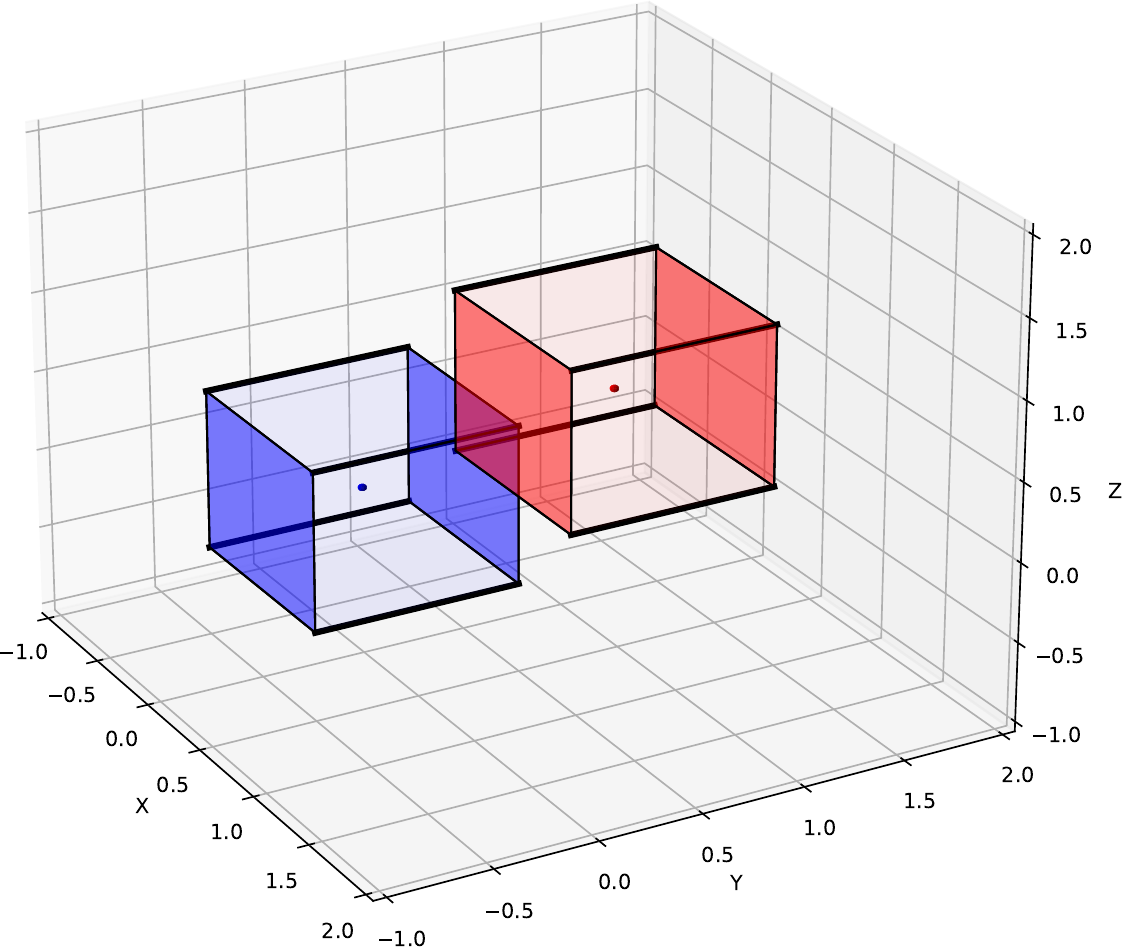}
  \end{subfigure}
  \hfill
  \begin{subfigure}[b]{0.3\linewidth}
    \centering
    \includegraphics[width=\textwidth, height=.2\textheight, page=1, keepaspectratio]{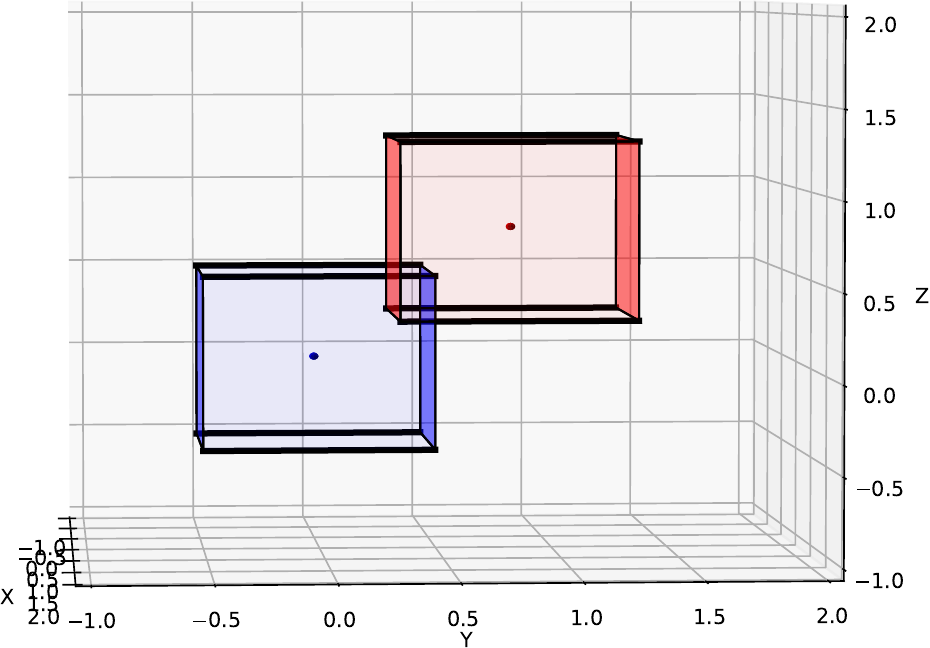}
  \end{subfigure}
  \hfill
  \begin{subfigure}[b]{0.3\linewidth}
    \centering
    \includegraphics[width=\textwidth, height=.2\textheight, page=1, keepaspectratio]{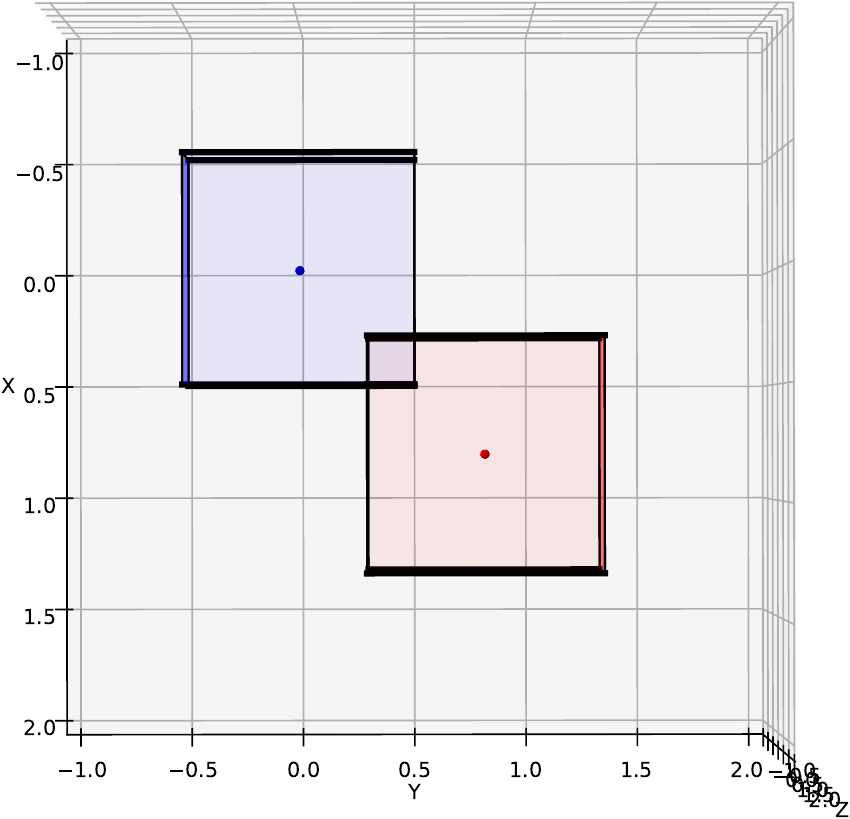}
  \end{subfigure}
  \caption{3D Rectangular representation of a source and target particles.
           Some rays (thick lines) intersect with the squares.}
  \label{fig:cube}
\end{figure}

From this description, one particle can be seen as a box of sides $(C + \epsilon , C + \epsilon , C)$ (see Figure~\ref{fig:cube}).
The rays can be seen as the four edges of the box in the $x$ direction, and the triangles composed the front and back faces.
If any two boxes have an intersection, we will detect it as shown in Figure~\ref{fig:triangular_representation}.

Potentially, the ray will intersect with the squares of the target particle, but this can easily be filtered by checking either the coordinates or the index of the geometric elements. 
Additionally, if the target and source particles are aligned on the $y$ or $z$ axis, two rays will intersect with the source's squares. 
To filter these intersections, we proceed as shown in Algorithm~\ref{alg:triangle_intersection}. 
We compare the coordinates between the source and the target and proceed as follows:
If $y$ and $z$ are different, we perform the computation (only the current ray will intersect).
If $y$ is equal, we use the ray of index 0 if $z$ is smaller, and the ray of index 2 if $z$ is greater.
If $z$ is equal, we use the ray of index 0 if $y$ is smaller, and the ray of index 1 if $y$ is greater.
Otherwise, only the ray of index 0 will be used for computation (all four rays will intersect).

\begin{algorithm}[htb!]
  \DontPrintSemicolon
  \SetKwFunction{optixGetPrimitiveIndex}{optixGetPrimitiveIndex}
  \SetKwFunction{optixGetGASTraversableHandle}{optixGetGASTraversableHandle}
  \SetKwFunction{optixGetSbtGASIndex}{optixGetSbtGASIndex}
  \SetKwFunction{optixGetTriangleVertexData}{optixGetTriangleVertexData}
  \SetKwFunction{getPayloadC}{getPayloadC}
  \SetKwFunction{getPayloadPartPos}{getPayloadPartPos}
  \SetKwFunction{distance}{distance}
  \SetKwFunction{getPayloadRayidx}{getPayloadRayidx}
  
  \KwData{Optix variables}
  \KwResult{Callback when a ray intersect with a triangle}
  \textbf{Function} callback()\;
  \Begin{       
    \tcc{Get information on the intersected triangle} 
    $vertices \gets$ \optixGetTriangleVertexData($gas, prim\_idx, sbtGASIndex, 0.f, vertices$)\;
    
    \tcc{Retreive the source position from the triangle vertices} 
    Declare $q$ as float3\;
    $q.y \gets (\max(vertices[0].y, vertices[1].y, vertices[2].y) + \min(vertices[0].y, vertices[1].y, vertices[2].y))/2$\;
    $q.z \gets (\max(vertices[0].z, vertices[1].z, vertices[2].z) + \min(vertices[0].z, vertices[1].z, vertices[2].z))/2$\;
    
    $c \gets \getPayloadC()$\;
    \If{$(prim\_idx \bmod 4) < 2$}{
        $q.x \gets vertices[0].x + c/2$\;
    }
    \Else{
        $q.x \gets vertices[0].x - c/2$\;
    }
    
    \tcc{Get target particle position} 
    $point \gets \getPayloadPartPos()$\;
    \tcc{Compute distance} 
    $dist\_p1\_p2 \gets \distance(point, q)$\;
    
    \tcc{Ensure it is not a self intersection and it is in the cutoff distance} 
    \If{$dist\_p1\_p2 < c AND dist\_p1\_p2 > \epsilon$}{
        $ray\_idx \gets \getPayloadRayidx()$\;
        $is\_ray\_for\_compute \gets (point.y \neq q.y$ AND $point.z \neq q.z)$ \;
        OR $((point.z < q.z$ AND $ray\_idx == 0)$ OR $(point.z > q.z$ AND $ray\_idx == 2)) $\;
        OR $((point.y < q.y$ AND $ray\_idx == 0)$ OR $(point.y > q.y$ AND $ray\_idx == 1)) $\;
        OR $ray\_idx == 0$\;
        
        \If{$is\_ray\_for\_compute$}{
          \tcc{Perform the computation} 
        }
    }
  }
  \caption{Triangle intersection callback}
  \label{alg:triangle_intersection}
\end{algorithm}

%%%%%%%%%%%%%%%%%%%%%%%%%%%%%%%%%%%%%%%%%%%%%%%%%%%%

\subsection{Custom AABB}
\label{sec:customaabb}

\begin{figure}[htb!]
  \centering
  \begin{subfigure}[b]{0.48\linewidth}
    \centering
    \includegraphics[width=\textwidth, height=.2\textheight, page=1, keepaspectratio]{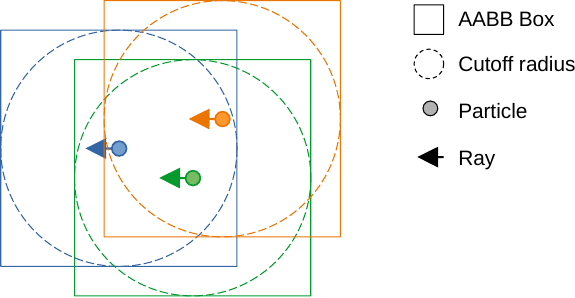}
  \end{subfigure}
  \hfill
  \begin{subfigure}[b]{0.48\linewidth}
    \centering
    \includegraphics[width=\textwidth, height=.15\textheight, page=1, keepaspectratio]{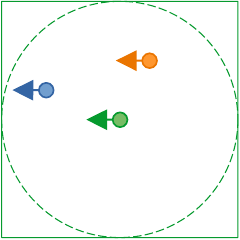}
  \end{subfigure}
\caption{Custom AABB representation of the particles.
         The rays are cast from the particles' positions with a infinitesimal length.
         In the IS callback invoked if a ray enters an AABB, the distance is checked and the interaction between the source and target particles is computed.
         Self interactions are filtered by ensuring that the source and target are different.}
\label{fig:aabboptix}
\end{figure}

In this strategy, a bounding box with a width equal to $2 \times C$ is created around each particle. 
OptiX treats these boxes as custom primitives, requiring us to implement a custom IS that is invoked when a ray enters an AABB. 
We do not perform any intersection tests within the IS, as our goal is not to compute ray-AABB or ray-surface intersections but to identify pairs of particles that are within a distance $C$ of each other.
To achieve this, rays are cast from the particles' positions with an infinitesimally short length, as illustrated in Figure~\ref{fig:aabboptix}.

The interaction between the source and target particles is computed directly within the IS callback.
Consequently, the hit and miss callbacks are left empty, as there is no need to filter intersections. 
In the IS callback, OptiX provides information about the ray (specifically, the starting point, which corresponds to the position of the source particle) and the index of the AABB being tested.
To obtain geometric information about the AABB, one could register a Primitive Geometry Acceleration Structure (PGAS) for each AABB during the OptiX scene build stage and retrieve this data in the IS callback. 
However, we observed that this approach significantly increased the scene build cost. 
Instead, we opted to access the global memory directly to retrieve the position of the target particle, which proved to be a more efficient solution.

%%%%%%%%%%%%%%%%%%%%%%%%%%%%%%%%%%%%%%%%%%%%%%%%%%%%%%%%%%%%%%%%%%%%%%%%%%%%%%%%%
%%%%%%%%%%%%%%%%%%%%%%%%%%%%%%%%%%%%%%%%%%%%%%%%%%%%%%%%%%%%%%%%%%%%%%%%%%%%%%%%%

\section{Performance study}
\label{sec:performance_study}

This section presents a performance evaluation of the different methods under various configurations. The analysis begins with a description of the experimental setup, including the hardware and software used for implementation. Performance is then examined for two types of particle distributions: uniform and non-uniform. Differentiating these distributions allows for assessing the efficiency of grid-based methods like CUDA, which are optimized for uniform distributions but can encounter inefficiencies with sparse or highly localized data in non-uniform scenarios.

%%%%%%%%%%%%%%%%%%%%%%%%%%%%%%%%%%%%%

\subsection{Experimental Setup.}

\paragraph{Hardware}
We have used two NVidia GPUs:
\begin{itemize}
\item A100~\footnote{\url{https://www.nvidia.com/content/dam/en-zz/Solutions/Data-Center/a100/pdf/nvidia-a100-datasheet-nvidia-us-2188504-web.pdf}} with 40GB hBM2, 48KB of shared-memory, 108 multi-processors, zero RT Cores, 8192 CUDA Cores max single-precision performance 19.5TFLOPS, and max tensor performance 311.84TFLOPS.
\item RTX8000~\footnote{\url{https://www.nvidia.com/content/dam/en-zz/Solutions/design-visualization/quadro-product-literature/quadro-rtx-8000-us-nvidia-946977-r1-web.pdf}} with 48GB GDDR6, 48KB of shared-memory, 72 multi-processors, 72 RT Cores, 4608 CUDA Cores, maximum Ray casting of 10Giga Rays/sec, max single-precision performance 16.3TFLOPS, and max tensor performance 130.5TFLOPS.
\end{itemize}

Despite the lack of RT cores, the A100 is capable of executing ray tracing kernels, the GPU then use its other units to behave similarly.

\paragraph{Software}

We have implemented the proposed approach in OptiX 8~\footnote{\url{https://developer.nvidia.com/rtx/ray-tracing/optix}}.
We use the GNU compiler 11.2.0 and the NVidia CUDA compiler 12.3.
The source code is available online~\footnote{\url{https://gitlab.inria.fr/bramas/particle-interaction-with-optix}}.

The code was compiled with the following flags: \texttt{-arch=sm\_75} for the RTX8000, \texttt{-arch=sm\_80} for the A100 (and \texttt{-O3 -DNDEBUG} on both).
We execute each kernel 5 times and take the average as reference.

\begin{figure}[htb!]
  \centering
  \begin{subfigure}[b]{0.49\linewidth}
      \centering
    \includegraphics[width=\textwidth, height=0.2\textheight, page=1, keepaspectratio]{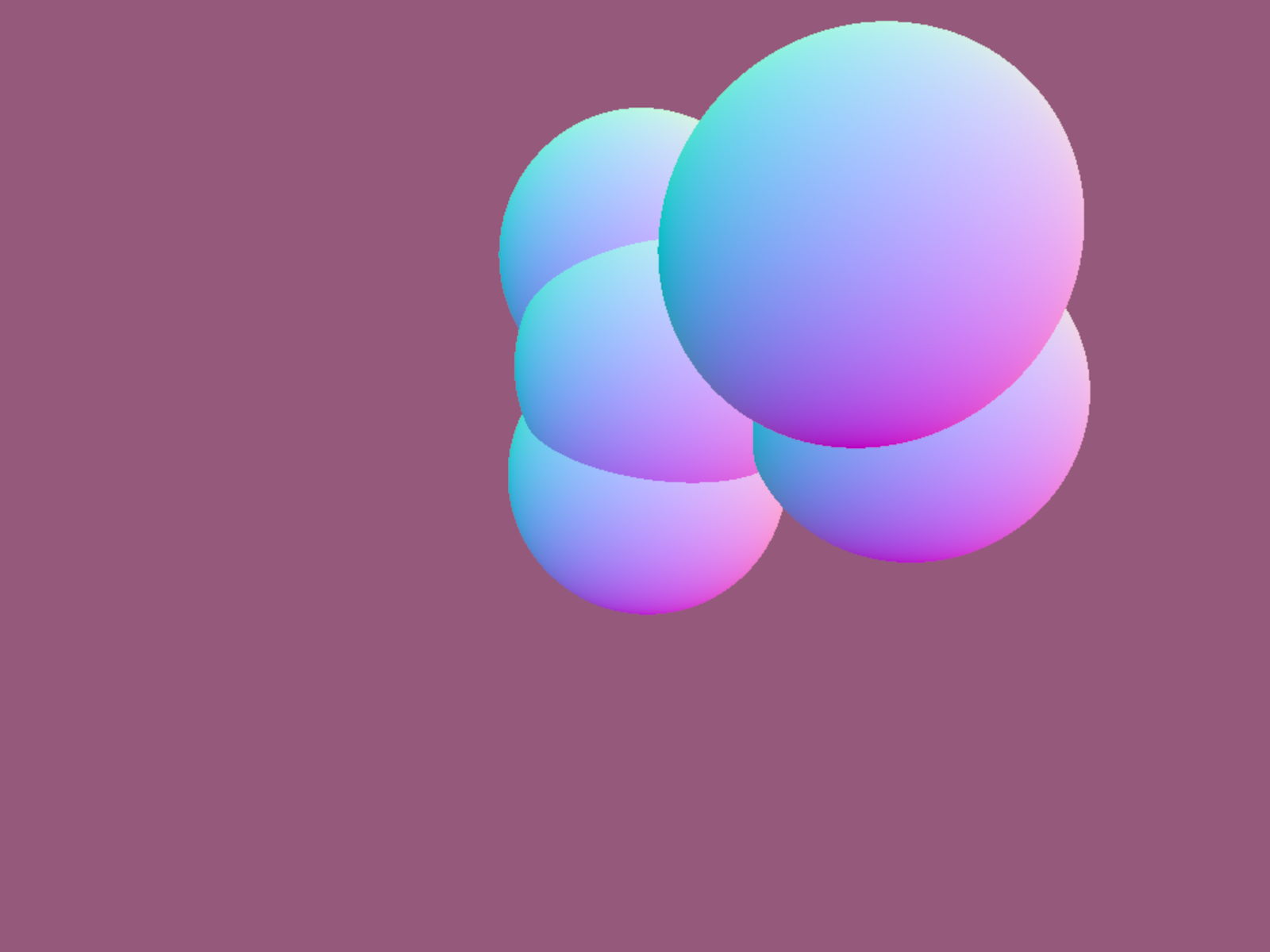}
    \caption{3D rendering of sphere primitives.}
    \label{fig:realspheres}
\end{subfigure}
\hfill
\begin{subfigure}[b]{0.49\linewidth}
    \centering
  \includegraphics[width=\textwidth, height=0.2\textheight, page=1, keepaspectratio, clip]{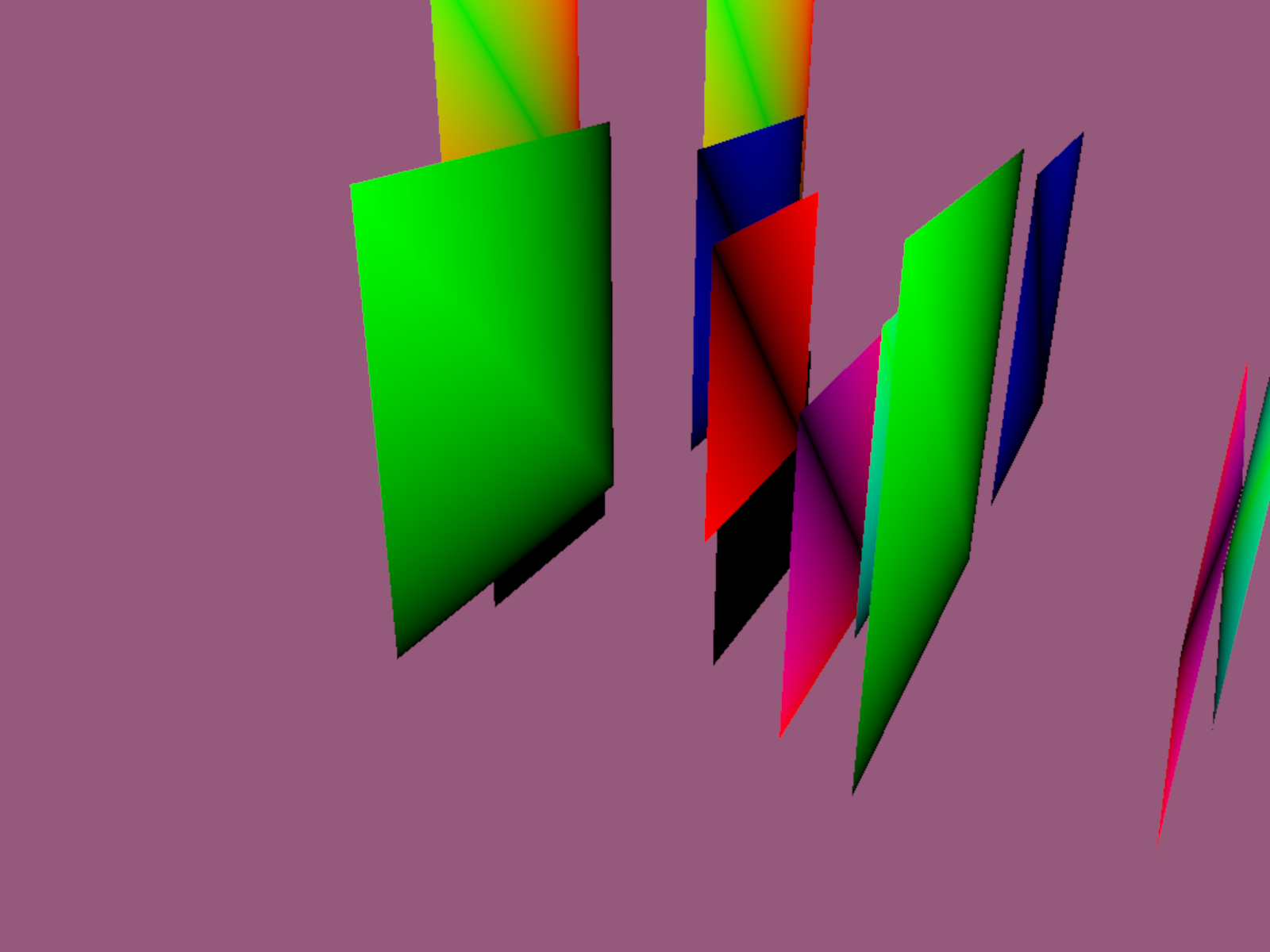}
  \caption{3D rendering of square primitives.}
  \label{fig:realtriangles}
\end{subfigure}
\caption{3D rendering of the primitives using conventional ray tracing.}
\label{fig:realprimitives}
\end{figure}

We provide in Figure~\ref{fig:realprimitives} the 3D rendering of the primitives using ray tracing: in Figure~\ref{fig:realspheres} for the spheres and in Figure~\ref{fig:realtriangles} for the squares.
These figures were drawn using the OptiX API and ray tracing from camera to the scene.

%%%%%%%%%%%%%%%%%%%%%%%%%%%%%%%%%%%%%
\subsection{Uniform distribution}

In this test case, the particles are distributed uniformly within a unit box. Consequently, there are no (or very few) empty cells in the grid of the CUDA version.
The cutoff distance is defined as $1 / \beta$, where $\beta$ can be $2$, $4$, $8$, $16$, or $32$. For a given $\beta$, the simulation grid consists of $3^\beta$ cells. The number of particles $N$ is then calculated as $N = p \times 3^\beta$, where $p$ represents the average number of particles per cell, taking values of $1$, $2$, $4$, $8$, $16$, or $32$. All computations are performed in single-precision floating point.

\begin{figure}[htp!]
  \centering
  \begin{subfigure}[b]{\linewidth}
      \centering
    \includegraphics[width=\textwidth, height=0.04\textheight, page=1, keepaspectratio]{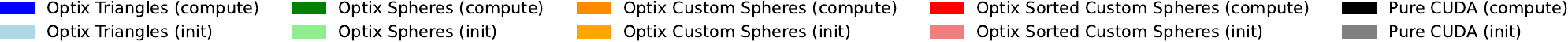}
\end{subfigure}

  \begin{subfigure}[b]{0.49\linewidth}
      \centering
    \includegraphics[width=\textwidth, height=0.2\textheight, page=1, keepaspectratio]{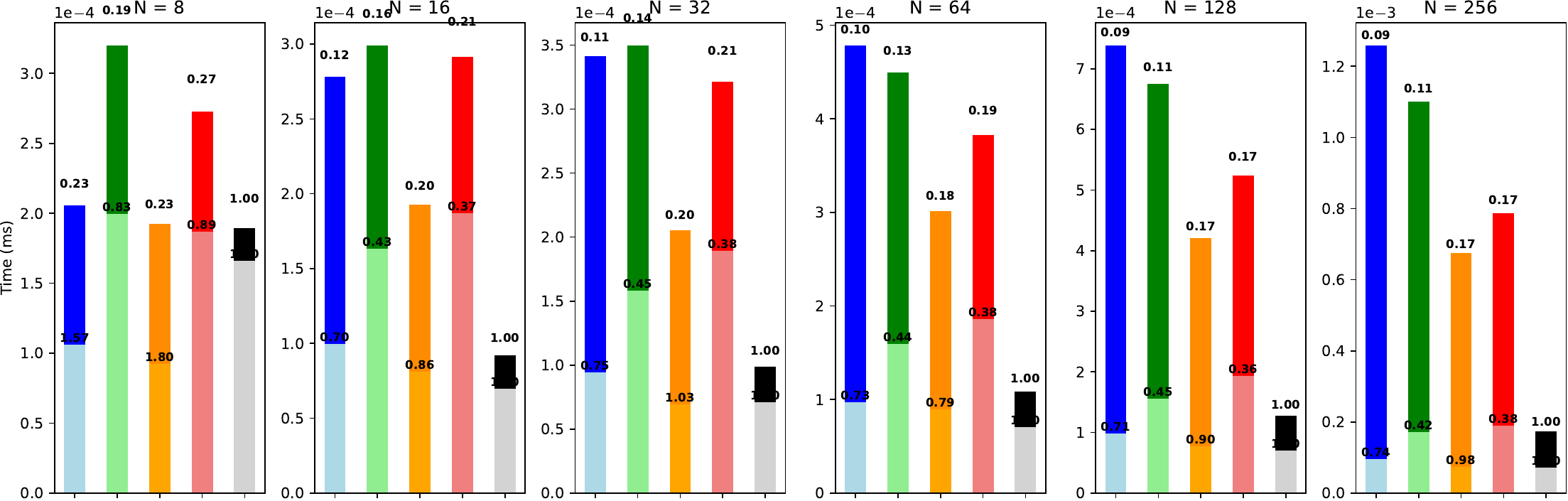}
    \caption{A100 $\beta =2$.}
    \label{fig:a100_2}
\end{subfigure}
\hfill
\begin{subfigure}[b]{0.49\linewidth}
    \centering
  \includegraphics[width=\textwidth, height=0.2\textheight, page=1, keepaspectratio]{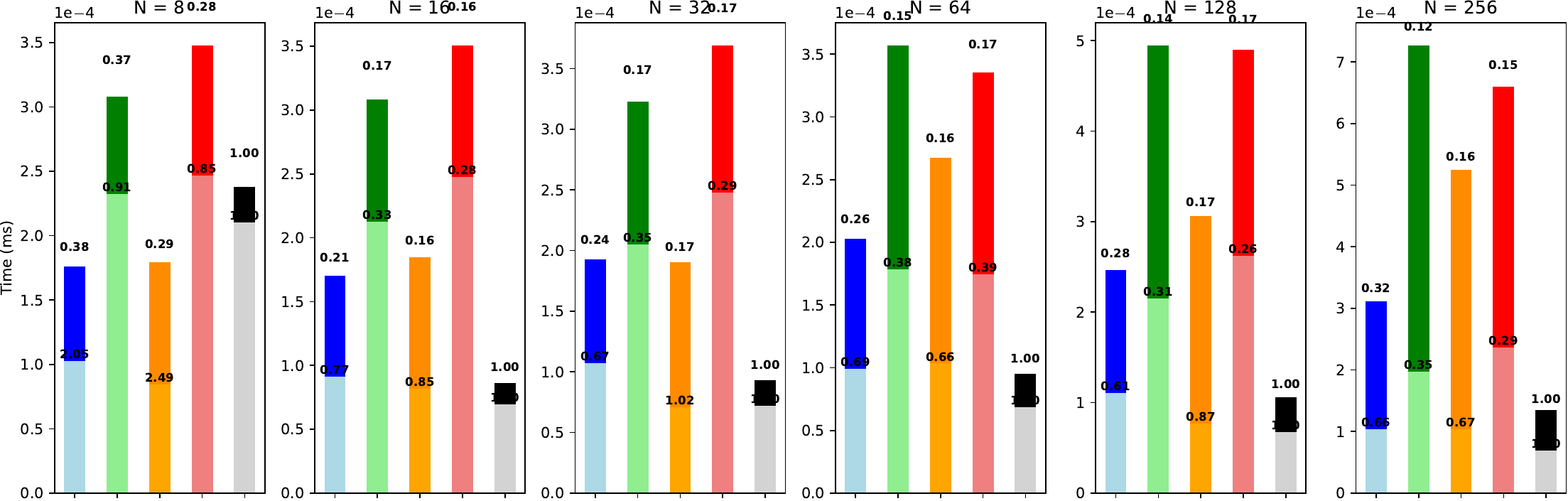}
  \caption{RTX8000 $\beta =2$.}
  \label{fig:rtx8000_2}
\end{subfigure}

  \begin{subfigure}[b]{0.49\linewidth}
      \centering
    \includegraphics[width=\textwidth, height=0.2\textheight, page=1, keepaspectratio]{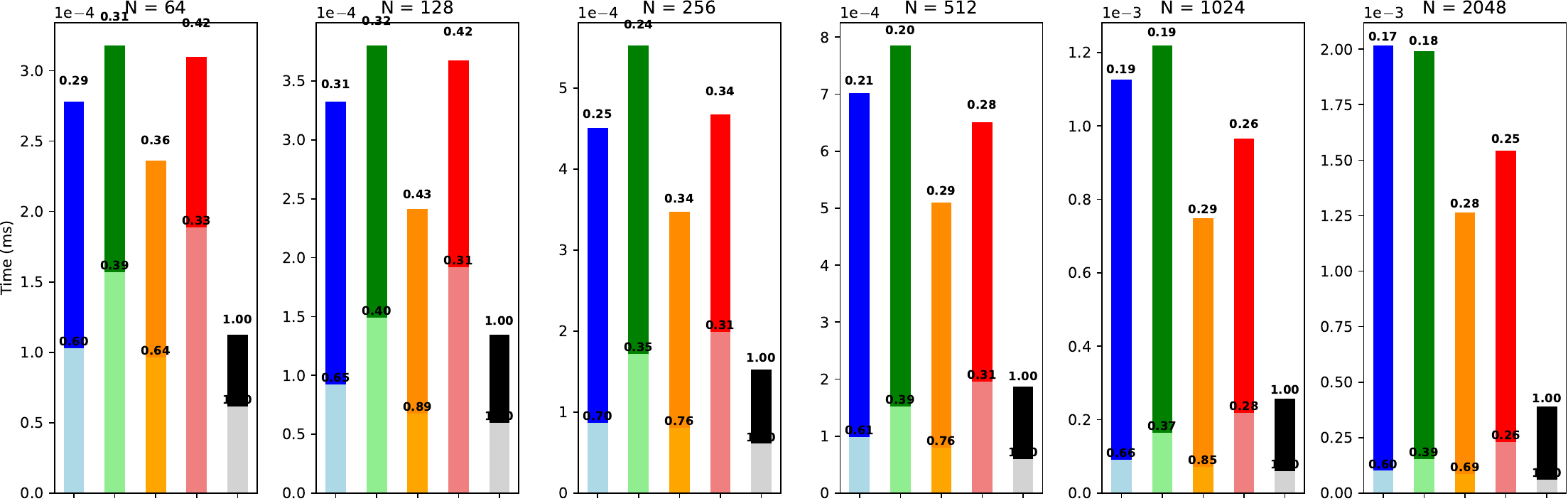}
    \caption{A100 $\beta =4$.}
    \label{fig:a100_4}
\end{subfigure}
\hfill
\begin{subfigure}[b]{0.49\linewidth}
    \centering
  \includegraphics[width=\textwidth, height=0.2\textheight, page=1, keepaspectratio]{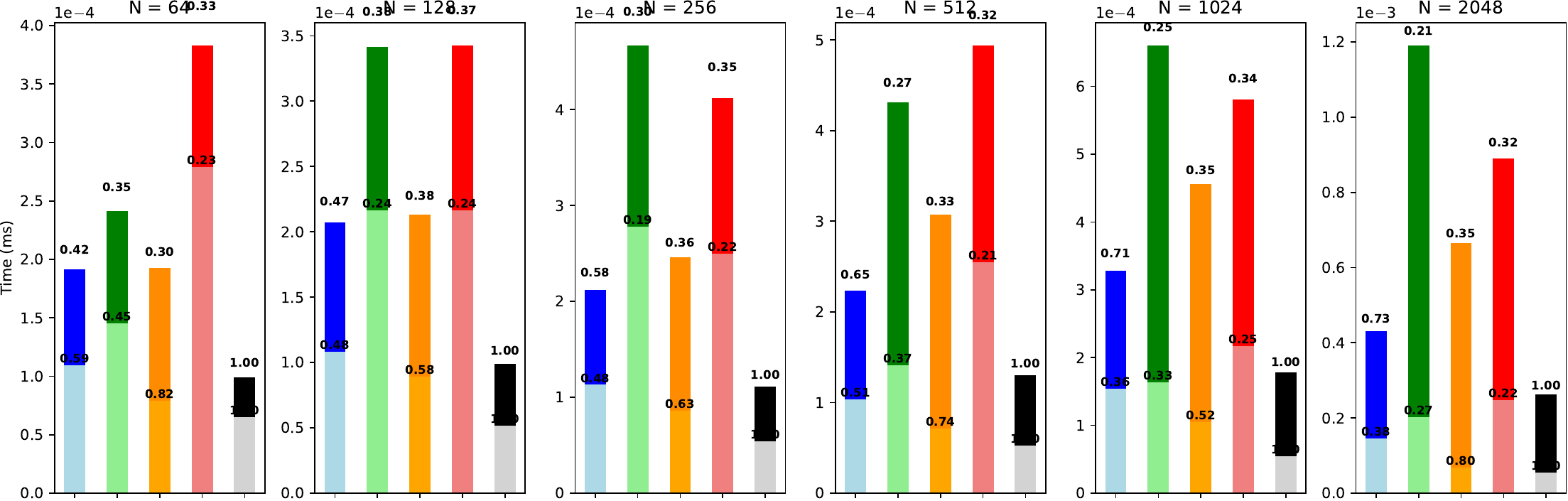}
  \caption{RTX8000 $\beta =4$.}
  \label{fig:rtx8000_4}
\end{subfigure}

  \begin{subfigure}[b]{0.49\linewidth}
      \centering
    \includegraphics[width=\textwidth, height=0.2\textheight, page=1, keepaspectratio]{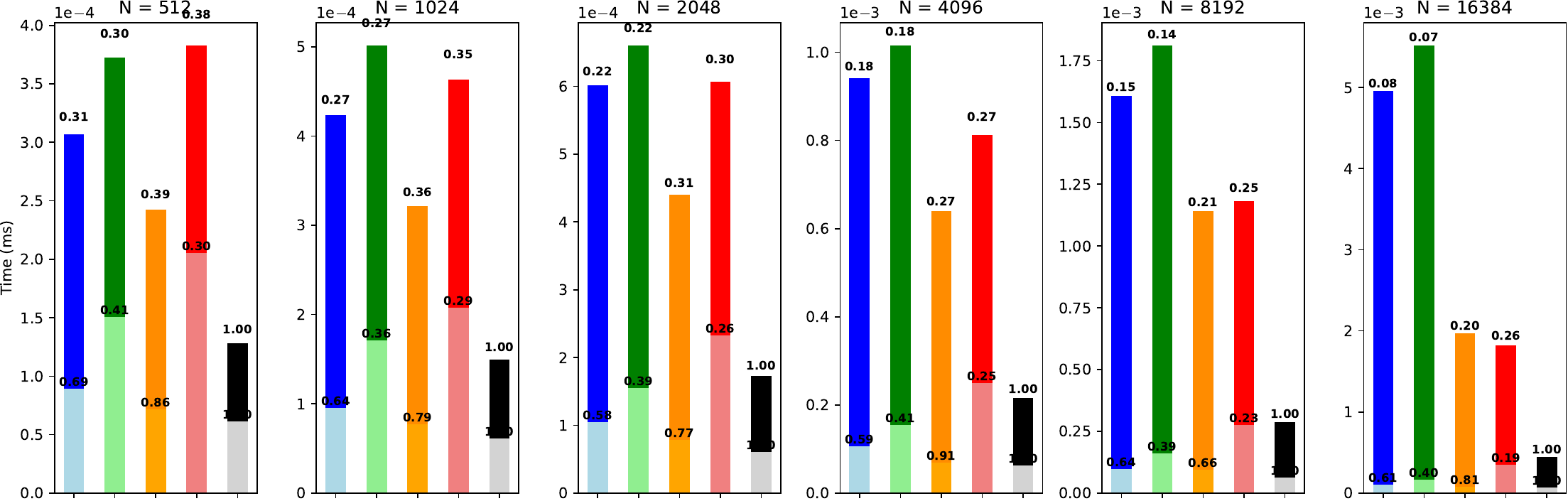}
    \caption{A100 $\beta =8$.}
    \label{fig:a100_8}
\end{subfigure}
\hfill
\begin{subfigure}[b]{0.49\linewidth}
    \centering
  \includegraphics[width=\textwidth, height=0.2\textheight, page=1, keepaspectratio]{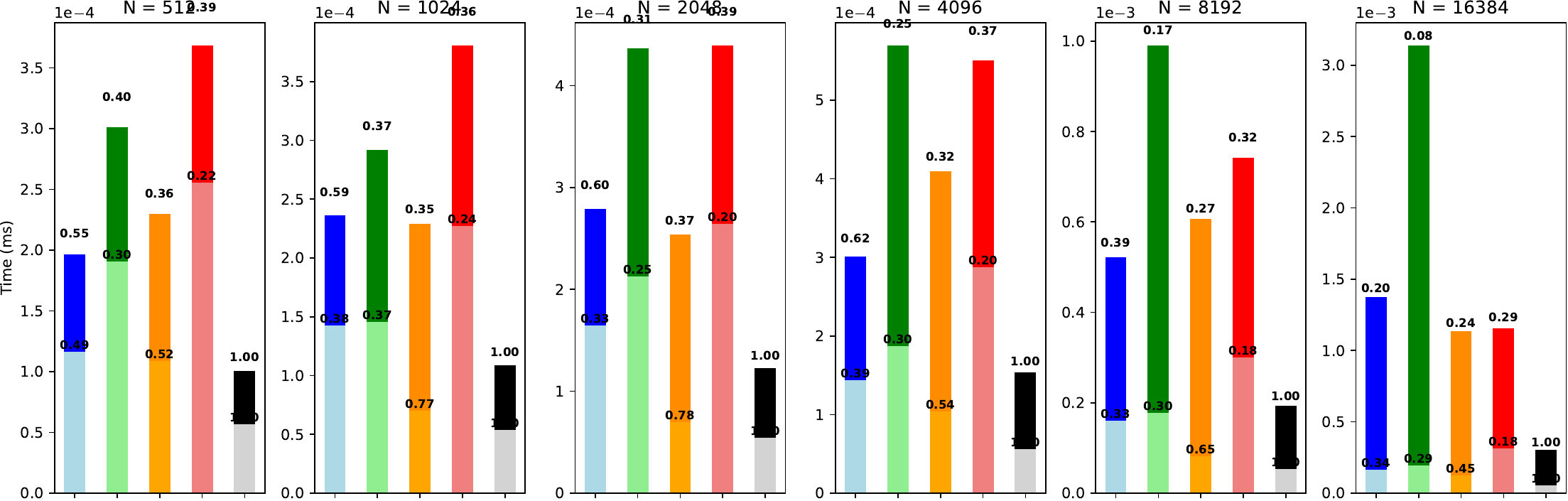}
  \caption{RTX8000 $\beta =8$.}
  \label{fig:rtx8000_8}
\end{subfigure}

  \begin{subfigure}[b]{0.49\linewidth}
      \centering
    \includegraphics[width=\textwidth, height=0.2\textheight, page=1, keepaspectratio]{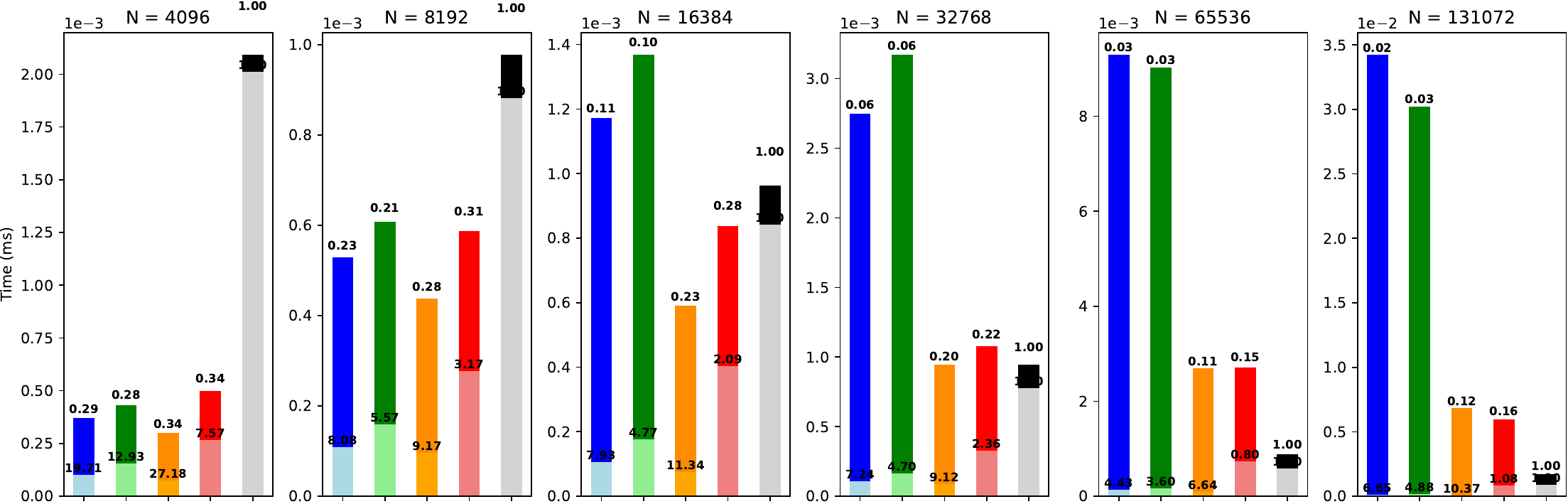}
    \caption{A100 $\beta =16$.}
    \label{fig:a100_16}
\end{subfigure}
\hfill
\begin{subfigure}[b]{0.49\linewidth}
    \centering
  \includegraphics[width=\textwidth, height=0.2\textheight, page=1, keepaspectratio]{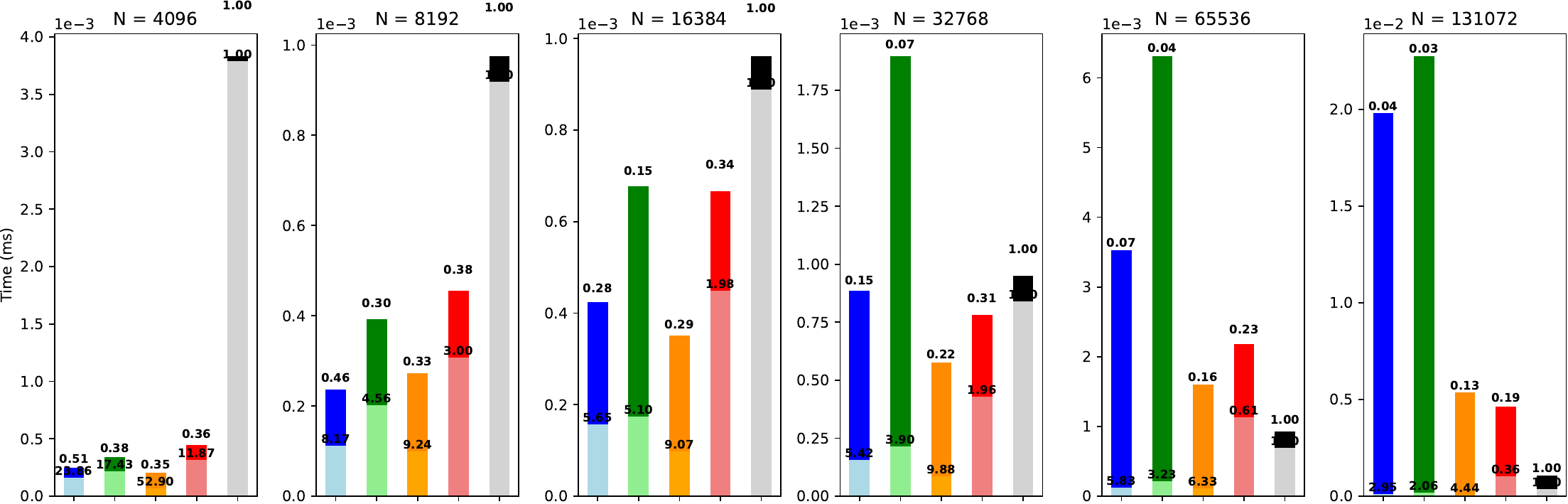}
  \caption{RTX8000 $\beta =16$.}
  \label{fig:rtx8000_16}
\end{subfigure}

  \begin{subfigure}[b]{0.49\linewidth}
      \centering
    \includegraphics[width=\textwidth, height=0.2\textheight, page=1, keepaspectratio]{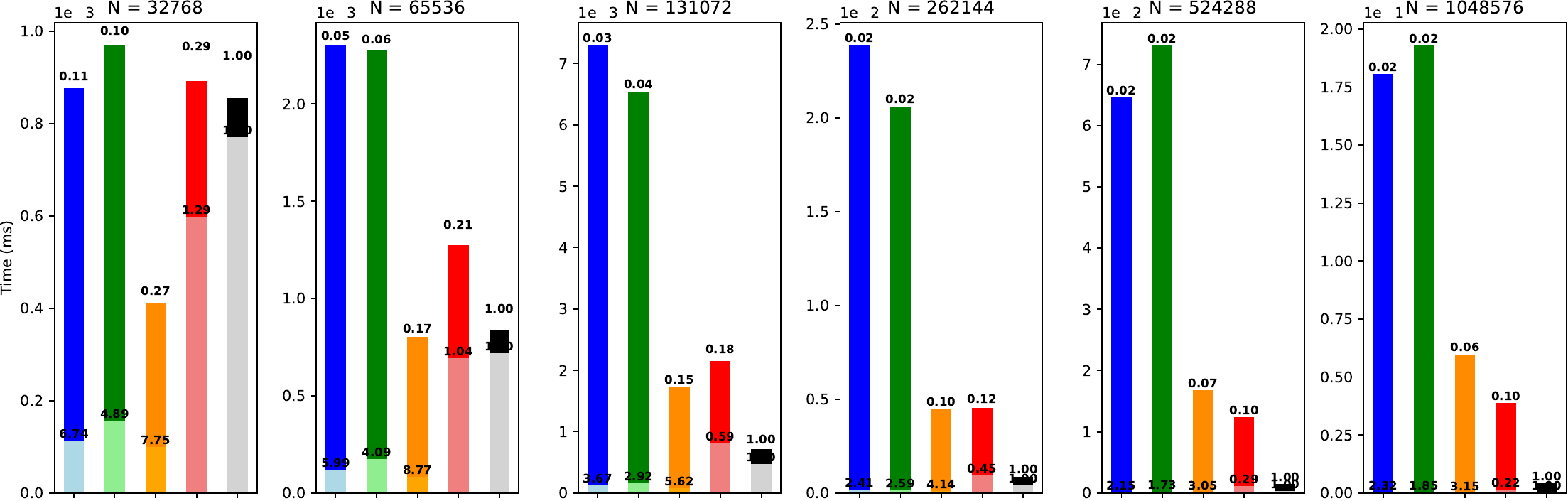}
    \caption{A100 $\beta =32$.}
    \label{fig:a100_32}
\end{subfigure}
\hfill
\begin{subfigure}[b]{0.49\linewidth}
    \centering
  \includegraphics[width=\textwidth, height=0.2\textheight, page=1, keepaspectratio]{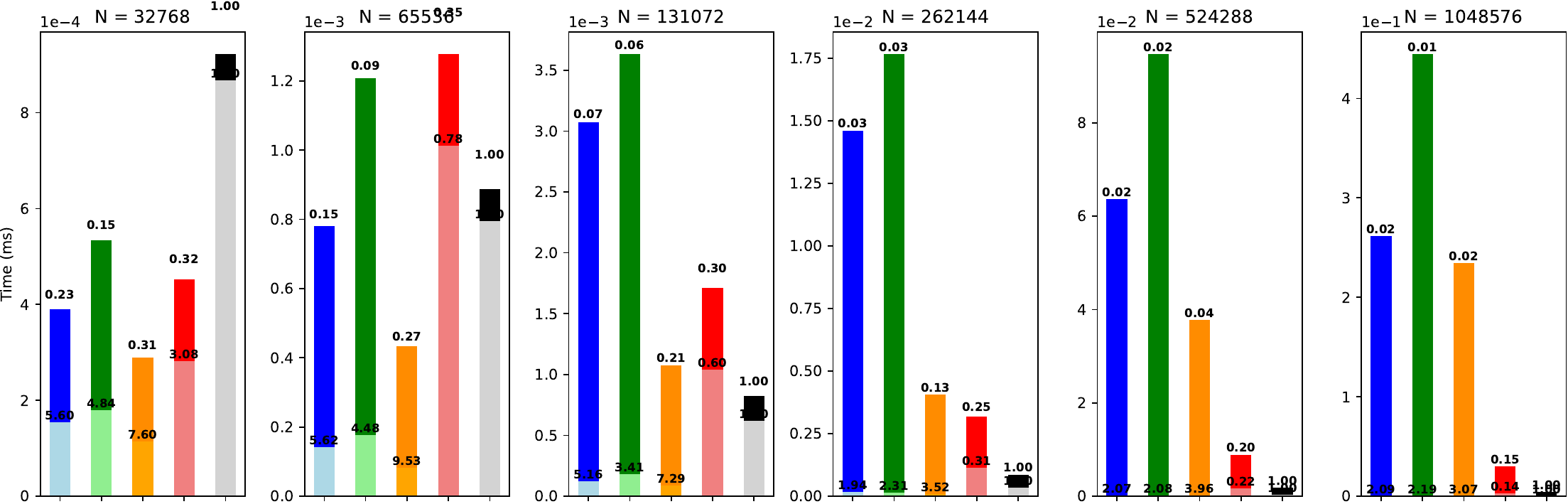}
  \caption{RTX8000 $\beta =32$.}
  \label{fig:rtx8000_32}
\end{subfigure}

\caption{Performance results for the two GPUs for our approaches (OptiX spheres, OptiX triangles, and Custom AABB without and with sort) and the pure CUDA implementation that use a grid of cells (CUDA).
         $\beta$ is the divisor coefficient of the simulation box.
         The cutoff distance is $1/\beta$ and there are $\beta^3$ cells in the grid in the Cuda version.
         The speedup against the CUDA version is shown above the bars for both the build and compute steps.}
\label{fig:perfresults}
\end{figure}

We present the results in Figure~\ref{fig:perfresults}. For all configurations, we measured the initialization step (light color) and the computation step (dark color). 

For the OptiX-based implementation, the initialization step involves building the scene by invoking the OptiX API to create the primitives. For the CUDA version, the initialization step involves constructing the grid of cells. Consequently, in the OptiX-based version, the computation step includes the time spent launching the rays, executing the callback functions, and performing the interactions. In the CUDA version, the computation step corresponds to the kernel time required to compute the interactions.

% We will plot 2 × 5 figures as sugfigures.

\paragraph{Comparison Between Triangles and Spheres}

First, we analyze the performance difference between triangles (blue) and spheres (green). On the A100 GPU, both models deliver similar performance, but the ratio of initialization time (light color) to computation time (dark color) is higher for spheres. This indicates that the computation step is more efficient for spheres than for triangles. This suggests that in scenarios where the initialization step is performed only once (e.g., static elements), spheres might be a better choice. However, on the RTX8000, triangles are faster than spheres across all configurations. Although the initialization step for spheres is quicker than for triangles in cases with fewer particles, this trend does not hold for larger configurations.

\paragraph{Comparison Between Custom AABB and Built-in Primitives}

Next, we compare the Custom AABB method (orange/red) with the built-in primitives (triangles in blue and spheres in green). On the A100, the Custom AABB consistently outperforms the built-in primitives, and its advantage grows as the number of particles or cells increases. For \(\beta = 32\) (Figure~\ref{fig:a100_32}), the Custom AABB is up to four times faster than the triangle- or sphere-based methods. On the RTX8000, the Custom AABB performs similarly to triangles for small particle counts but becomes faster as the number of particles increases. Overall, the Custom AABB is the fastest method. This demonstrates that creating a custom Intersection Shader (IS) that is significantly simpler and lighter than the built-in ones, which compute actual intersections, can accelerate execution.

\paragraph{Benefit of Sorting the Particles in the Custom AABB Methods}
Sorting the particles to position them closer in memory based on their spatial proximity in the simulation is expected to optimize memory accesses. However, sorting the particles also incurs a computational cost. We observe that the strategy without sorting (orange) is faster than the strategy with sorting (red) when there are few particles. However, as the number of particles increases, the benefits of sorting become significant. In our test case, for a given $\beta$, increasing the number of particles leads to an increase in neighboring particles and, consequently, the number of interactions. In such cases, optimizing memory access becomes critical.

\paragraph{GPU Performance Comparison}

We then compare the performance of the two GPUs. Results for the A100 are presented in Figures~\ref{fig:a100_2}, \ref{fig:a100_4}, \ref{fig:a100_8}, \ref{fig:a100_16}, and \ref{fig:a100_32}, while results for the RTX8000 are shown in Figures~\ref{fig:rtx8000_2}, \ref{fig:rtx8000_4}, \ref{fig:rtx8000_8}, \ref{fig:rtx8000_16}, and \ref{fig:rtx8000_32}. Both GPUs exhibit comparable performance and a similar initialization-to-computation time ratio. Although the A100 is expected to offer higher raw performance, the OptiX implementation is not fully optimized for this GPU, while the RTX8000 benefits from having more RT cores to accelerate computation. Additionally, we do not utilize the tensor cores of the A100, which represent a key differentiator between the two GPUs and could help approach theoretical performance.

\paragraph{Comparison Between OptiX and CUDA Implementations}

Finally, we compare the OptiX triangles (blue), OptiX spheres (green), Custom AABB (orange), Custom sorted AABB (red), and CUDA implementation (green). The OptiX-based implementation is generally slower than the CUDA implementation for $\beta$ values from $2$ to $8$, except in a few cases (e.g., $\beta=2$ and $N=8$). In these scenarios, the initialization step in the CUDA implementation is negligible, and the computation step dominates. Consequently, avoiding the grid of cells does not yield a significant advantage. For $\beta=8$ (Figures~\ref{fig:a100_8} and \ref{fig:rtx8000_8}), the CUDA computation step is shorter because the OptiX implementations include not only the interaction computation but also the intersection list computation using rays. Additionally, the memory access pattern in OptiX-based implementations differs, as spatially close particles in memory may not be close in space and may have varying neighbors.

For $\beta=16$ and $\beta=32$, the Custom AABB performs better for configurations with few particles per cell (e.g., one or two particles per cell on average in the CUDA version). In these cases, the computation cost is low, and the CUDA version spends most of its time in the initialization step. Furthermore, the CUDA version incurs overhead for empty cells because it constructs a dense grid. For such configurations, the OptiX-based implementations allocate more time to the computation step, making the Custom AABB a suitable choice when the number of particles per cell is low, particularly in scenarios where moving particles require frequent grid rebuilding.

%%%%%%%%%%%%%%%%%%%%%%%%%%%%%%%%%%%%%%%%%%%%%%%%%%%%%%%%%%%%%%%%%%%%%%%%%%%%%%%%%
%%%%%%%%%%%%%%%%%%%%%%%%%%%%%%%%%%%%%%%%%%%%%%%%%%%%%%%%%%%%%%%%%%%%%%%%%%%%%%%%%
\subsection{Non Uniform Distribution}

In this test case, we distribute the particles on the surface of a unit sphere. Consequently, most of the cells on the grid in the CUDA version are empty. With this aim, we use a coefficient $\alpha$, which can be $8$, $16$, $32$, $64$ or $128$. For a given $\alpha$, the simulation grid consists of approximately $3^\alpha$ non-empty cells. The number of particles $N$ is then calculated as $N = p \times 3^\alpha$, where $p$ represents the average number of particles per cell, taking values of $1$, $2$, $4$, $8$, $16$, or $32$. The cutoff radius is set such that each particle should have approximately $9 \times p$ neighbor particles (it will be in general smaller than $2/\alpha$). In order to facilitate the reproducibility of our test case, we provide in Appendix the Code~\ref{lst:gensurface} that generates the different configurations. All computations are performed in single-precision floating point.

\begin{figure}[htp!]
  \centering
  \begin{subfigure}[b]{\linewidth}
      \centering
    \includegraphics[width=\textwidth, height=0.04\textheight, page=1, keepaspectratio]{./legend_only_figure}
\end{subfigure}

  \begin{subfigure}[b]{0.49\linewidth}
      \centering
    \includegraphics[width=\textwidth, height=0.2\textheight, page=1, keepaspectratio]{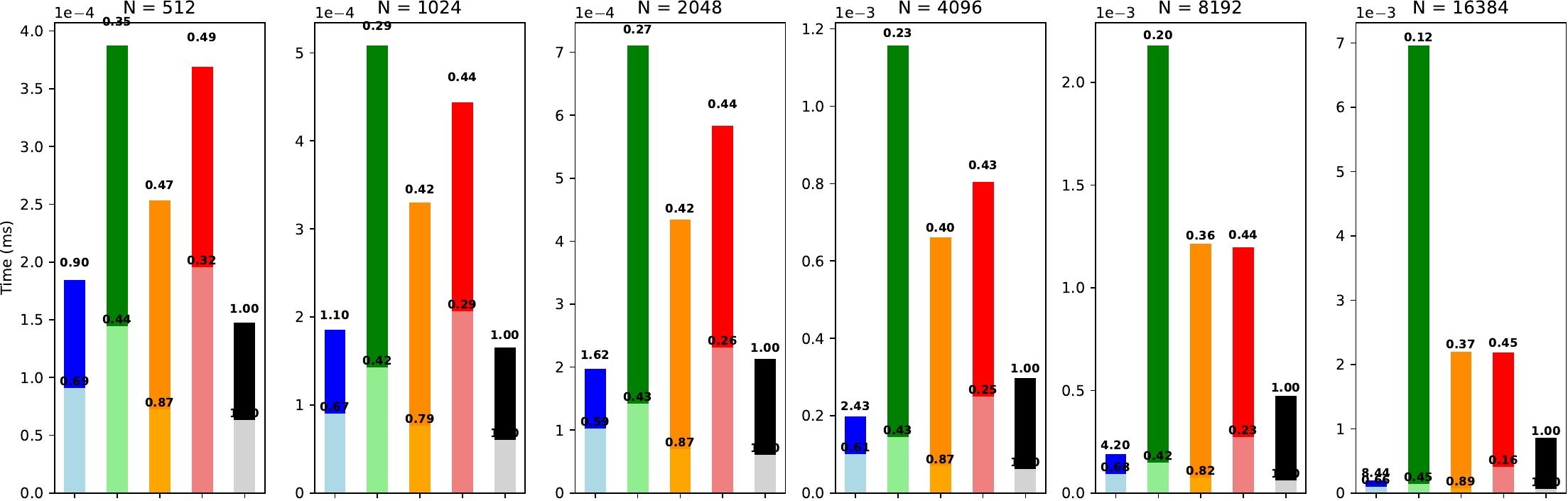}
    \caption{A100 $\alpha =8$.}
    \label{fig:a100_8_surface}
\end{subfigure}
\hfill
\begin{subfigure}[b]{0.49\linewidth}
    \centering
  \includegraphics[width=\textwidth, height=0.2\textheight, page=1, keepaspectratio]{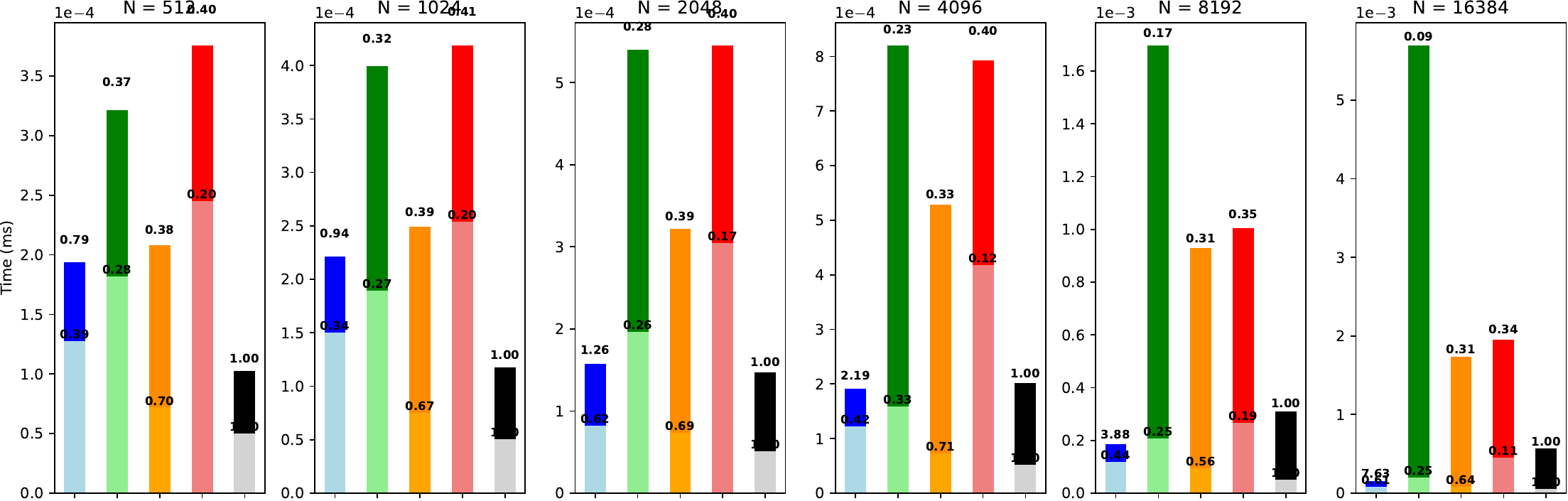}
  \caption{RTX8000 $\alpha =8$.}
  \label{fig:rtx8000_8_surface}
\end{subfigure}

  \begin{subfigure}[b]{0.49\linewidth}
      \centering
    \includegraphics[width=\textwidth, height=0.2\textheight, page=1, keepaspectratio]{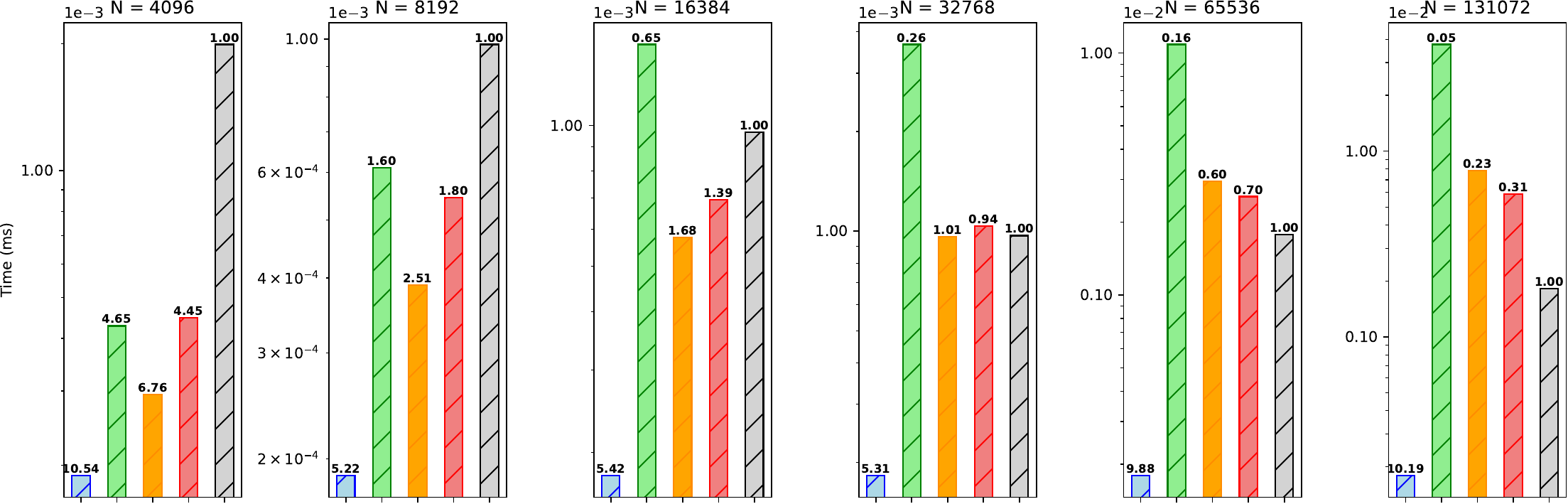}
    \caption{A100 $\alpha =16$ (Log.).}
    \label{fig:a100_4}
\end{subfigure}
\hfill
\begin{subfigure}[b]{0.49\linewidth}
    \centering
  \includegraphics[width=\textwidth, height=0.2\textheight, page=1, keepaspectratio]{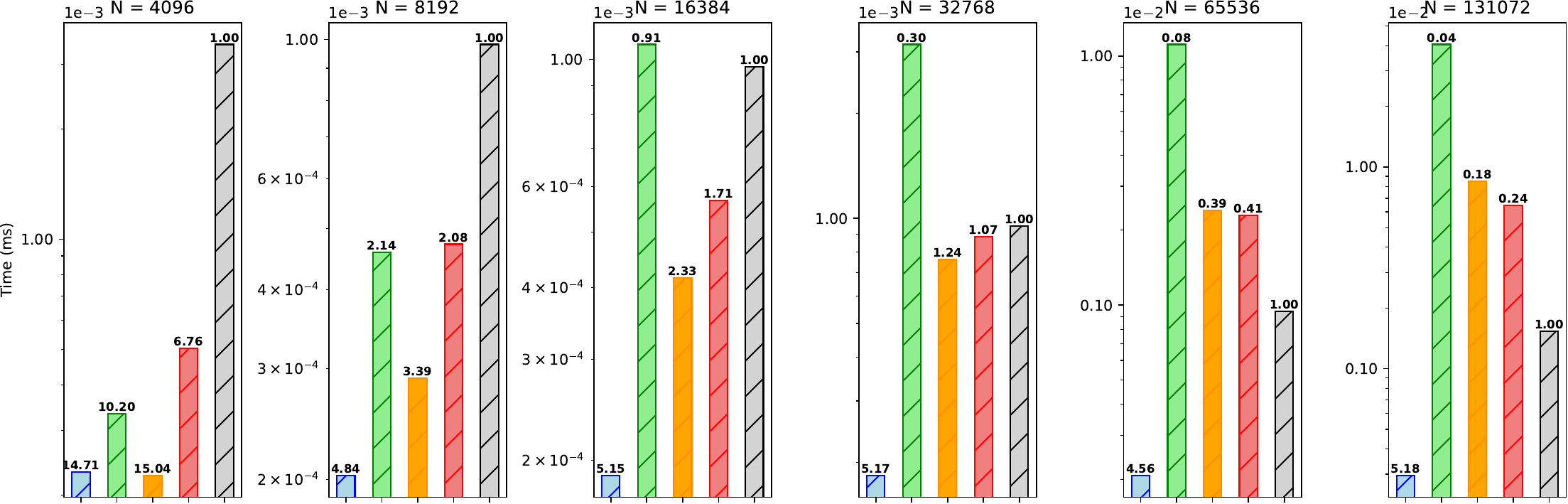}
  \caption{RTX8000 $\alpha =16$ (Log.).}
  \label{fig:rtx8000_16_surface_log}
\end{subfigure}

  \begin{subfigure}[b]{0.49\linewidth}
      \centering
    \includegraphics[width=\textwidth, height=0.2\textheight, page=1, keepaspectratio]{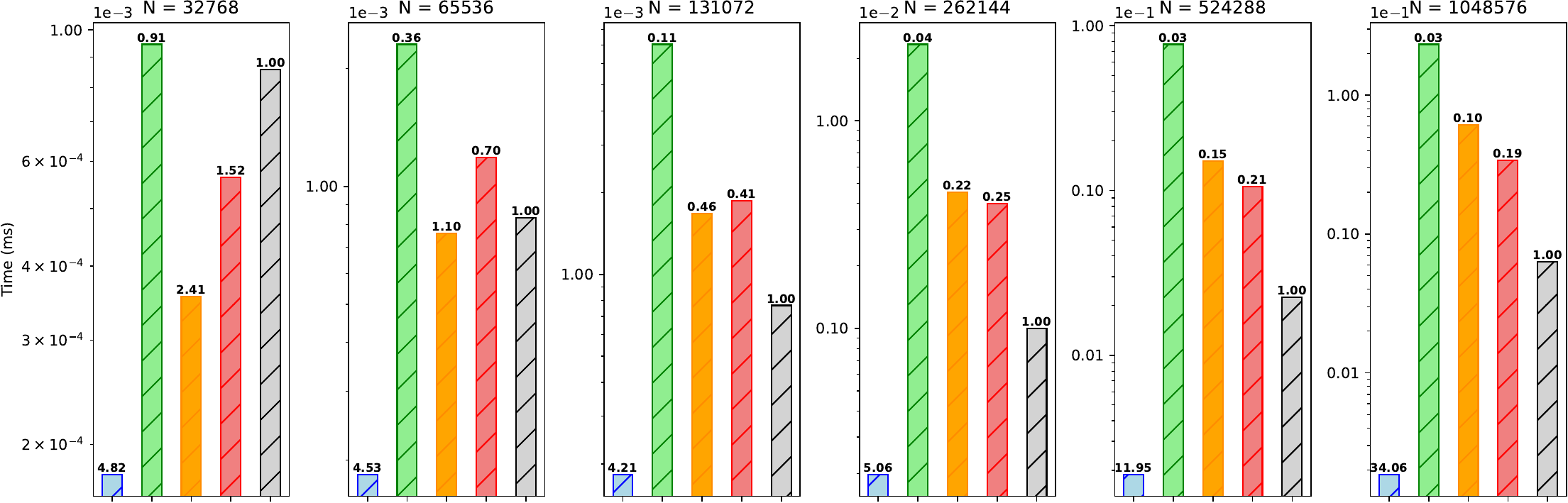}
    \caption{A100 $\alpha =32$ (Log.).}
    \label{fig:a100_32_surface_log}
\end{subfigure}
\hfill
\begin{subfigure}[b]{0.49\linewidth}
    \centering
  \includegraphics[width=\textwidth, height=0.2\textheight, page=1, keepaspectratio]{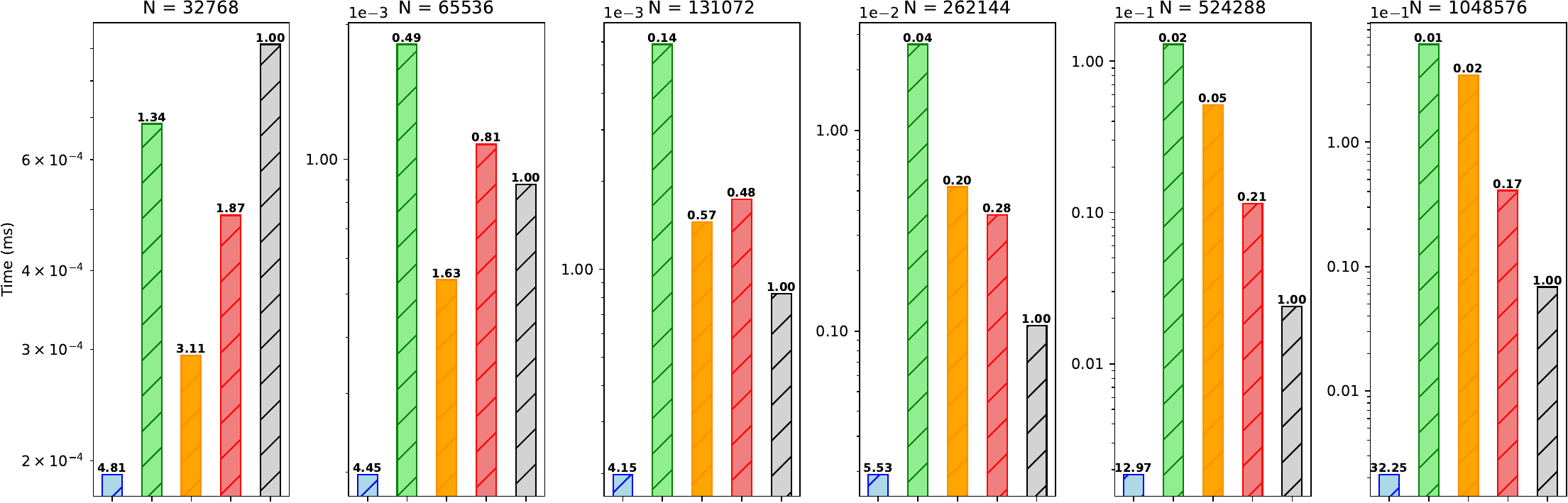}
  \caption{RTX8000 $\alpha =32$ (Log.).}
  \label{fig:rtx8000_32_surface_log}
\end{subfigure}

  \begin{subfigure}[b]{0.49\linewidth}
      \centering
    \includegraphics[width=\textwidth, height=0.2\textheight, page=1, keepaspectratio]{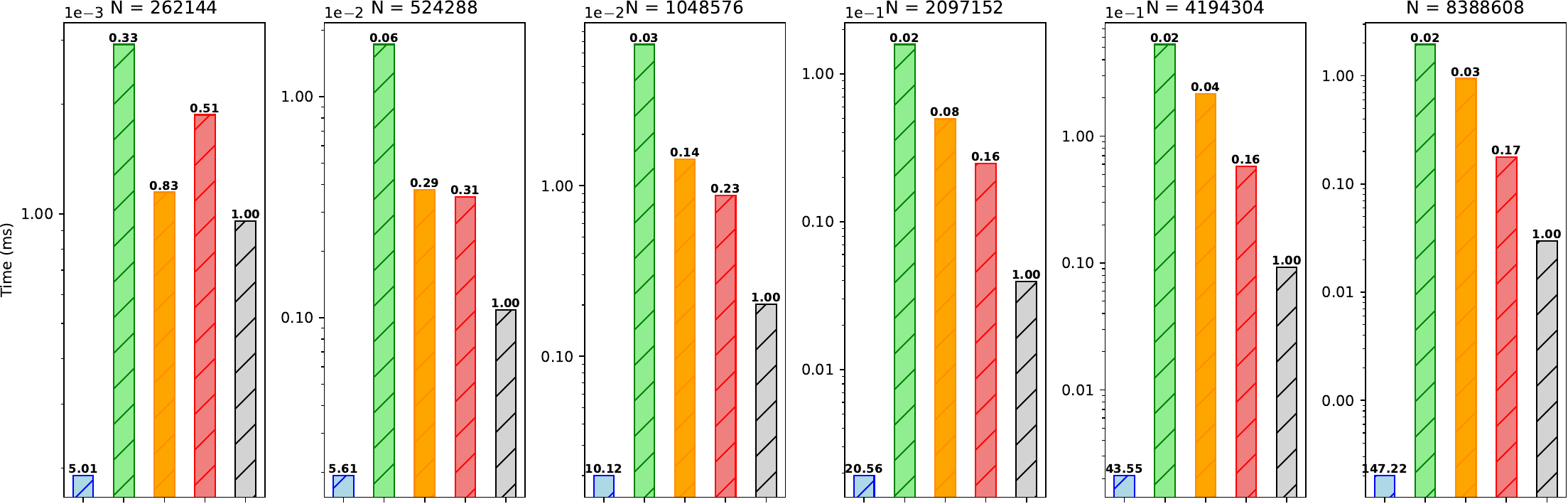}
    \caption{A100 $\alpha =64$ (Log.).}
    \label{fig:a100_64_surface_log}
\end{subfigure}
\hfill
\begin{subfigure}[b]{0.49\linewidth}
    \centering
  \includegraphics[width=\textwidth, height=0.2\textheight, page=1, keepaspectratio]{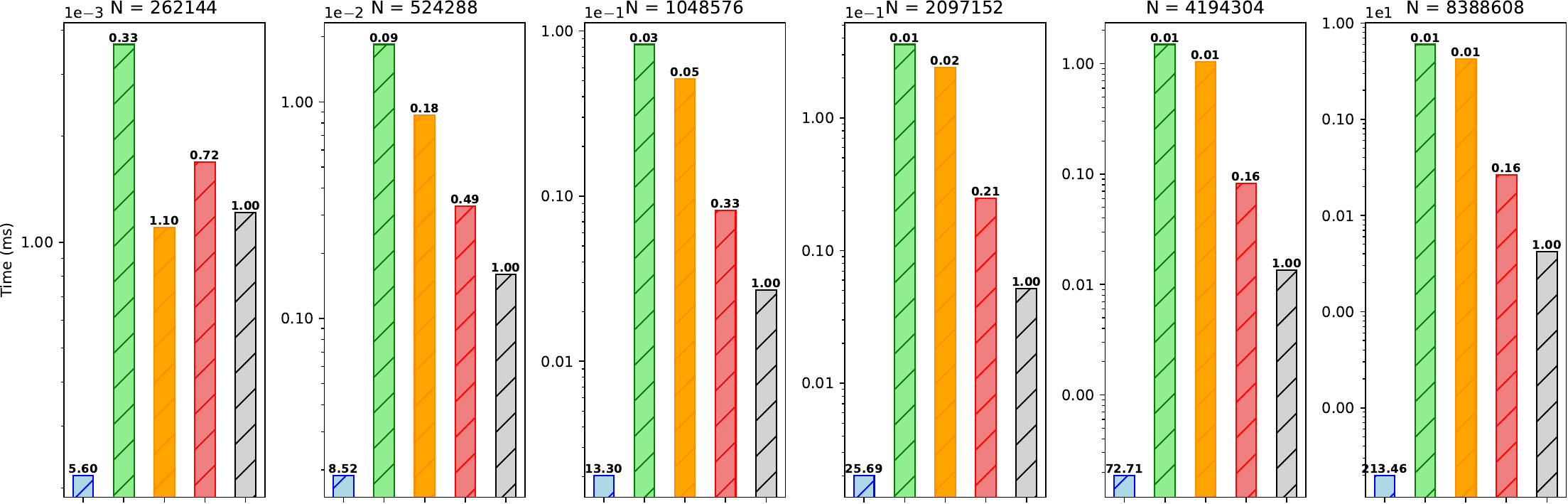}
  \caption{RTX8000 $\alpha =64$ (Log.).}
  \label{fig:rtx8000_64_surface_log}
\end{subfigure}

  \begin{subfigure}[b]{0.49\linewidth}
      \centering
    \includegraphics[width=\textwidth, height=0.2\textheight, page=1, keepaspectratio]{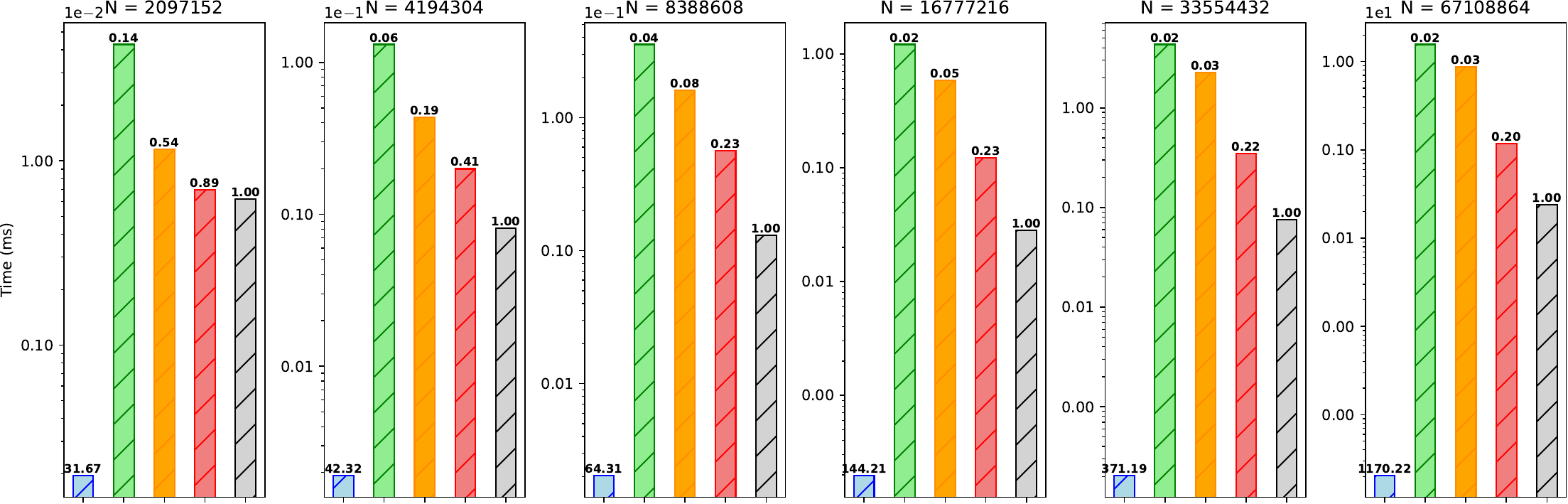}
    \caption{A100 $\alpha =128$ (Log.).}
    \label{fig:a100_128_surface_log}
\end{subfigure}
\hfill
\begin{subfigure}[b]{0.49\linewidth}
    \centering
  \includegraphics[width=\textwidth, height=0.2\textheight, page=1, keepaspectratio]{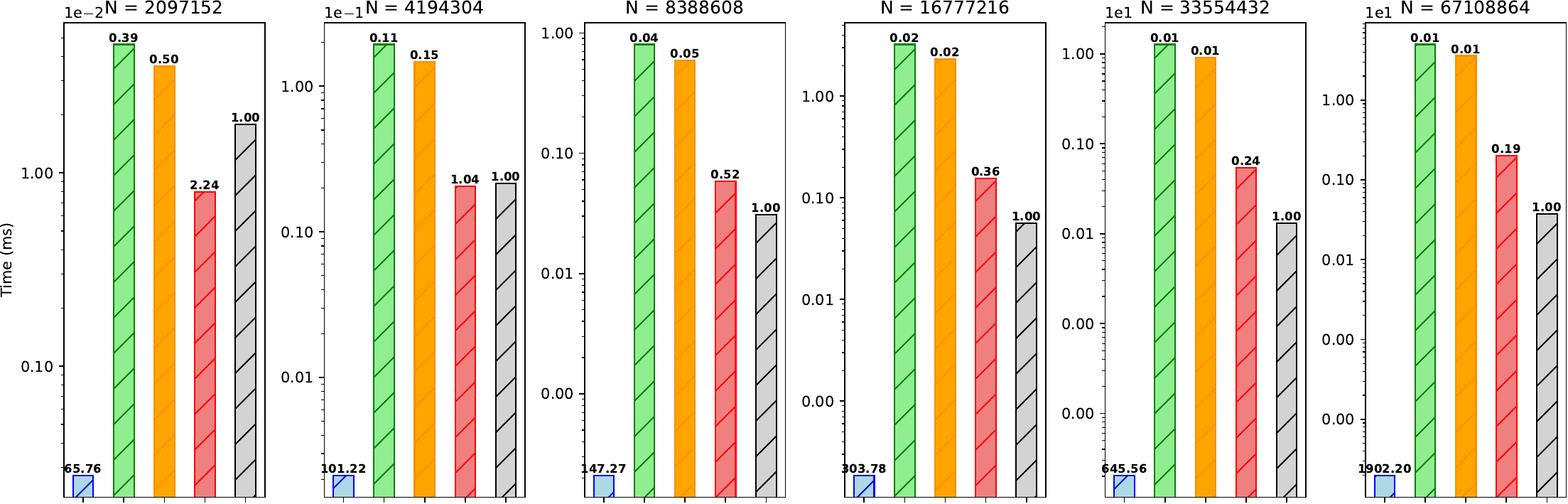}
  \caption{RTX8000 $\alpha =128$ (Log.).}
  \label{fig:rtx8000_128_surface_log}
\end{subfigure}

\caption{Performance results for the two GPUs for our approaches (OptiX spheres, OptiX triangles, and Custom AABB without and with sort) and the pure CUDA implementation that use a grid of cells (CUDA).
         $\alpha$ is the divisor coefficient of the simulation box.
         The cutoff distance is lower than $1/\beta$ and there are around $\alpha^3$ non-empty cells in the grid in the Cuda version.
         The speedup against the CUDA version is shown above the bars for both the build and compute steps, or for the complete run.}
\label{fig:perfresultssurface}
\end{figure}

We present the results in Figure~\ref{fig:perfresultssurface}. For all configurations, we measured both the initialization step (light color) and the computation step (dark color) for the first configuration. However, for the others, we directly present the total execution time using a logarithmic scale on the y-axis due to extreme differences.

For the OptiX-based implementation, the initialization step involves building the scene by invoking the OptiX API to create the primitives. For the CUDA version, the initialization step involves constructing the grid of cells. Consequently, in the OptiX-based version, the computation step includes the time spent launching the rays, executing the callback functions, and performing the interactions. In the CUDA version, the computation step corresponds to the kernel time required to compute the interactions.

% We will plot 2 × 5 figures as sugfigures.

\paragraph{OptiX Spheres} The OptiX spheres method is consistently the slowest (except for $\alpha = 2$ and a small number of particles). It has a more expensive initialization step compared to the Custom AABB methods and a significantly larger computation step. This is because the built-in IS kernel for spheres is computationally intensive, as it calculates several values, such as the norm of the intersection, that we do not use since we are not rendering an image.

\paragraph{Custom AABB Without and With Sorting} For low $\alpha$, sorting does not provide an advantage; it increases the cost of the initialization step too much relative to the gains in the computation step. However, for high $\alpha$ and a large number of particles, sorting the particles results in a performance improvement. This demonstrates that random access to global memory is so costly that the overhead of sorting the particles is worthwhile. The Custom AABB method with sorting performs comparably to the CUDA version. However, the Custom AABB methods are generally slower than both the CUDA version and the OptiX triangles method.

\paragraph{OptiX Triangles} This strategy is the most efficient. For $\alpha = 8$, its initialization step is comparable in cost to the OptiX spheres and Custom AABB methods. However, its computation step is significantly faster. We attribute this to two primary reasons. First, the built-in IS kernel for triangles is likely much simpler than that for spheres and is probably heavily optimized by NVIDIA, as triangles are the most commonly used primitive in 3D rendering. Second, the AABB bounding boxes around triangles are much smaller than those around spheres or the custom AABB primitives. This enables faster tree traversal and fewer false positives (i.e., cases where a ray intersects a bounding box but not the primitive inside it).
Finally, the CUDA version includes several components in its implementation that rely on algorithms with complexity linear in the number of cells. Since the vast majority of these cells are empty, these steps become highly inefficient. This is particularly evident for $\alpha = 8$, where the initialization step is notably prominent. Moreover, if we were to create a test case with $\alpha = 256$ (not included in the current study), the GPU would run out of memory for the the CUDA version allocating an excessively large grid. This limitation makes alternative approaches based on a tree structure — such as our OptiX-based methods — the only viable options.

%%%%%%%%%%%%%%%%%%%%%%%%%%%%%%%%%%%%%%%%%%%%%%%%%%%%%%%%%%%%%%%%%%%%%%%%%%%%%%%%%
%%%%%%%%%%%%%%%%%%%%%%%%%%%%%%%%%%%%%%%%%%%%%%%%%%%%%%%%%%%%%%%%%%%%%%%%%%%%%%%%%

\section{Conclusion}
\label{sec:conclusion}

In this paper, we proposed leveraging ray tracing technology to compute particle interactions within a cutoff distance in a 3D environment. We introduced one method that uses custom primitives and a custom Intersection Shader (IS), similar to the state-of-the-art, and two methods that use built-in primitives and actual intersection computations. For the latter, we described geometric algorithms to build the interaction list based on ray intersections with spheres or triangles.

Our results indicate that our approach provides a modest advantage in the preprocessing stage by avoiding the construction of a grid of cells. In addition, it is slower than the classical approach during the computation step when dealing with large numbers of uniformly distributed particles. However, when most cells are empty, our approach can provide a significant speedup. Therefore, we believe these methods have the potential to deliver better performance in the future or for specific applications, and we hope our work inspires the community to explore this direction further.

As GPU architectures continue to evolve, with advancements such as NVIDIA's Ada Lovelace architecture featuring third-generation RT cores and AMD's RDNA 3 architecture incorporating second-generation ray-tracing accelerators, we anticipate that algorithms leveraging these capabilities will become increasingly effective and relevant. Consequently, our approach is prepared to benefit from these hardware improvements, enhancing its performance and applicability in future computational scenarios and we hope our work inspires the community to explore this direction further.

We plan to evaluate our approach on other GPUs, such as AMD Radeon, and aim to port our method even in cases where the IS cannot be customized, thanks to our geometric algorithms.

%%%%%%%%%%%%%%%%%%%%%%%%%%%%%%%%%%%%%%%%%%%%%%%%%%%%%%%%%%%%%%%%%%%%%%%%%%%%%%%%%
%%%%%%%%%%%%%%%%%%%%%%%%%%%%%%%%%%%%%%%%%%%%%%%%%%%%%%%%%%%%%%%%%%%%%%%%%%%%%%%%%
\section*{Acknowledgments}

Experiments presented in this paper were carried out using the PlaFRIM experimental test-bed, supported by Inria, 
CNRS (LABRI and IMB), Université de Bordeaux, Bordeaux INP and Conseil Régional d’Aquitaine (see https://www.plafrim.fr).
    
%%%%%%%%%%%%%%%%%%%%%%%%%%%%%%%%%%%%%%%%%%%%%%%%%%%%%%%%%%%%%%%%%%%%%%%%%%%%%%%%%
%%%%%%%%%%%%%%%%%%%%%%%%%%%%%%%%%%%%%%%%%%%%%%%%%%%%%%%%%%%%%%%%%%%%%%%%%%%%%%%%%

% \bibliography{main}

%%%%%%%%%%%%%%%%%%%%%%%%%%%%%%%%%%%%%%%%%%%%%%%%%%%%%%%%%%%%%%%%%%%%%%%%%%%%%%%%%
%%%%%%%%%%%%%%%%%%%%%%%%%%%%%%%%%%%%%%%%%%%%%%%%%%%%%%%%%%%%%%%%%%%%%%%%%%%%%%%%%

\section*{Appendix}

\section{Discussion on the Sphere Model}
\label{app:proof}

The filtering algorithm presented in Section \ref{sec:spherical_representation} is based on the assumption that there is always at least one ray that is intersected once by the sphere and that it is the closest one to the center of the sphere. The following section is dedicated to demonstrating this hypothesis.

\begin{figure}[htb!]
  \centering
  \includegraphics[width=\textwidth, page=3, keepaspectratio, clip, trim=0cm 6cm 6cm 0cm]{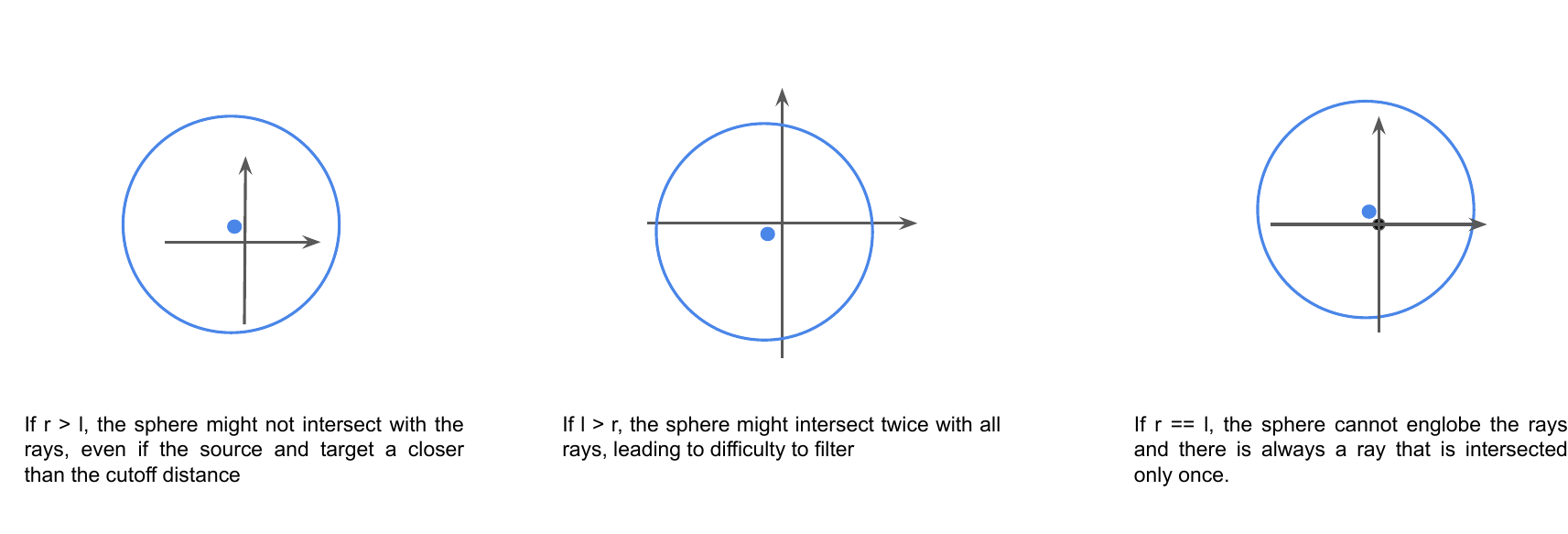}
  \caption{2D Spherical representation of the particles in three different cases with $r > l$, $l > r$ and $r == l$.}
  \label{fig:spherical_lr}
  \end{figure}

In Figure~\ref{fig:spherical_lr}, we show the different possibilities depending on the radius $r$ and ray's length $l$.
As it can be seen, even if the sources located at a distance of $C$ from the target could have their spheres that intersect with the rays when $r > l$,
we must set $l = r$ to ensure that the rays will intersect with the sphere in all cases, especially when the source and target are close. 

For the clarity of the proof, a demonstration is first provided in the plane, followed by a generalization to $\mathbb{R}^3$.
%%%%%%%%%%%%%%%%%%%%%%%%%%%%%%%%%%%%%%%%%%%%
\paragraph{2D Case}
Let us consider a cercle of radius $r$ centered at $(x_c, y_c)$, and a cross~\footnote{That is, two orthogonal axis-aligned segments of the same size intersecting at their respective centers.} centered at the origin with a length of $r$.
We consider the cases where the cercle is located in the first quarter, i.e., $ 0 \leq x_s \leq r$ and $ 0 \leq y_s \leq r$, but the demonstration remains valid for other quarters.

The equation of a cercle is given by:
\begin{equation}
(x - x_c)^2 + (y - y_c)^2 = r^2.
\end{equation}

The coordinates of the intersection points of the cercle with the axis are given by:
\begin{equation}
  \begin{split}
x_0 &= x_c - \sqrt{r^2 - y_c^2} \quad \text{and} \quad x_1 = x_c + \sqrt{r^2 - y_c^2} \\
y_0 &= y_c - \sqrt{r^2 - x_c^2}. \quad \text{and} \quad y_1 = y_c + \sqrt{r^2 - x_c^2}.
\end{split}
\end{equation}

\begin{figure}[htb!]
  \centering
  \includegraphics[width=\textwidth, page=4, keepaspectratio, clip, trim=0cm 8cm 0cm 0cm]{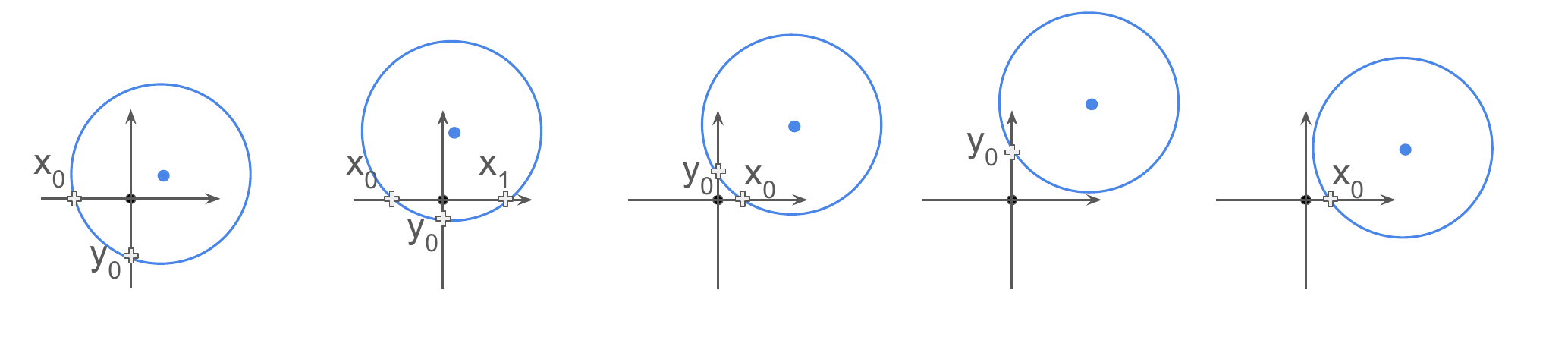}
  \caption{2D spherical representation illustrating the classification of intersection points between the sphere and the cross.}
  \label{fig:spherical_cross}
  \end{figure}

We provide the Figure~\ref{fig:spherical_cross} that shows where these points are located on the sphere and the cross.

$x_0$ and $y_0$ are the coordinates of the intersection points of the cercle that remain on the cross as the cercle get away from the origin.
On the other hand, $x_1$ and $y_1$ are the coordinates of the intersection points of the cercle that are the farthest from the origin of the cross and that can potentially be too far to remain on the cross (they will be on the corresponding axis but behind $l$).

\begin{lemma}
  Given a circle $\mathcal{C}$ and let $e_x$ and $e_y$ be the two segments of the cross of length less than $r$, then the number of intersections of $\mathcal{C}$ with $e_x$ or with $e_y$ is strictly less than $2$.
\end{lemma}

\textbf{Proof:}
As we consider that it is impossible that both $x_1$ and $y_1$ exist at the same time, we consider that these two equations cannot be true at the same time
\begin{equation}
x_c + \sqrt{r^2 - y_c^2} \leq r \quad \text{and} \quad y_c + \sqrt{r^2 - x_c^2} \leq r.
\end{equation}
Which can be simplified as
\begin{equation}
  \begin{split}
x_c < r - \sqrt{r^2-y_c^2} \quad \text{and} \quad & \sqrt{r^2 - x_c^2} \leq r - y_c \\
& r^2 - x_c^2 \leq r^2 - 2 r y_c + y_c^2 \\
& x_c^2 > 2 r y_c - y_c^2 \\
& x_c > \sqrt{2 r y_c - y_c^2},
\end{split}
\end{equation}
ending up with
\begin{equation}
  \sqrt{2 r y_c - y_c^2} < r - \sqrt{r^2-y_c^2}.
\end{equation}

\begin{figure}[htb!]
  \centering
  \includegraphics[width=\linewidth, height=.2\textheight, keepaspectratio]{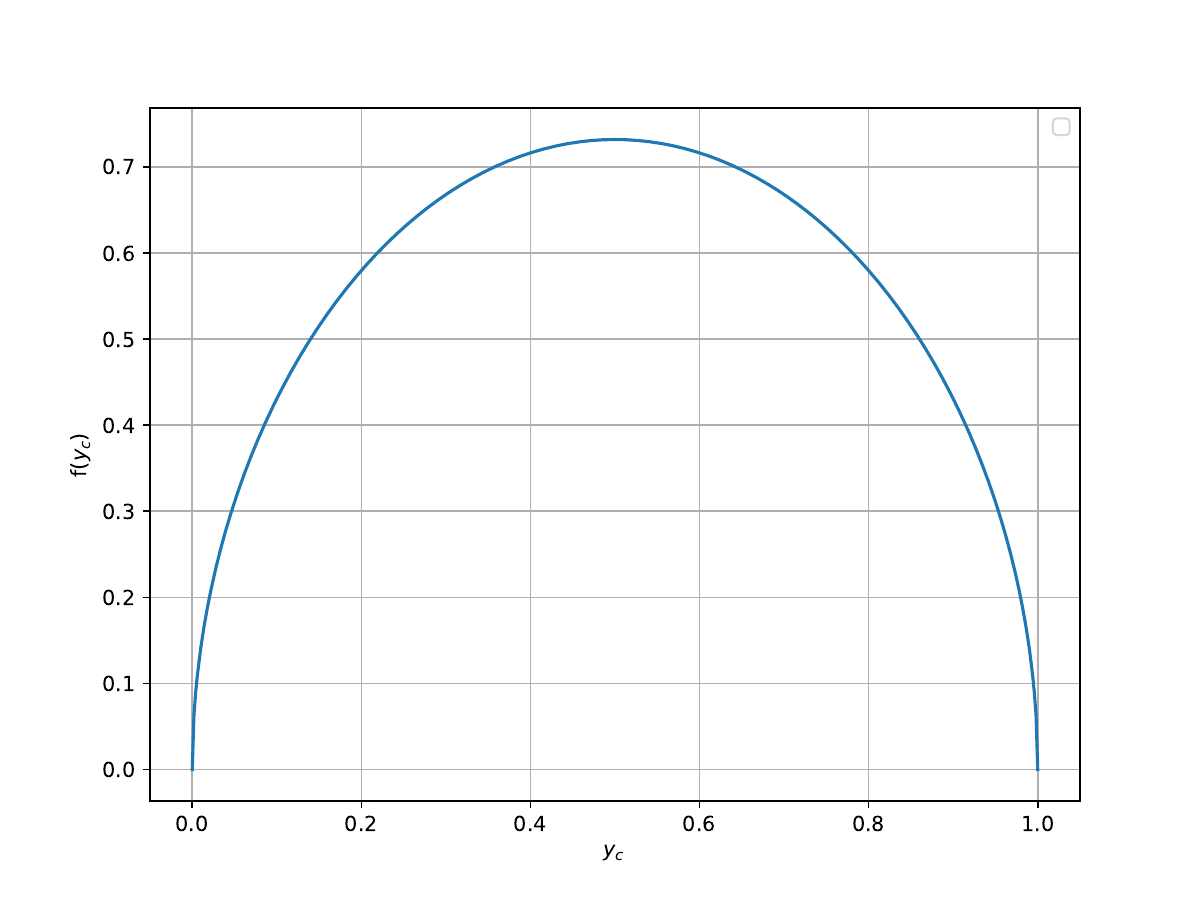}
  \caption{Plot of the equation $\sqrt{2ry_c - y_c^2} - r + \sqrt{r^2 - y_c^2}$, for r = 1.}
  \label{fig:limiteformula}
\end{figure}

For our definition range of $y_c \in [0, r]$, this equation has no solution (see Figure~\ref{fig:limiteformula}), which confirms that $\mathcal{C}$ will always intersect with the cross at most once for $e_x$ or $e_y$.

%%%%%%%%%%%%%%%%%%%%%%%%%%%%%%%%%%%%%%%%%%%%

The second statement is to show that the ray that is intersected once is the closest to the center of the sphere.
\begin{lemma}
  Given a circle $\mathcal{C}$ and $e$ one of the two segments of the cross of length less than $r$, such as the number of intersection of $\mathcal{C}$ with $e$ is equal to $1$,
   then the distance of $e$ with $\mathcal{C}$ is smaller than the distance of the other bar to $\mathcal{C}$.
\end{lemma}

\textbf{Proof:}
Consider that $\mathcal{C}$ intersects with $e_y$ twice and $e_x$ once, it means that $y_1 < r$ and $x_1 > r$, i.e. $y_c + \sqrt{r^2 - x_c^2} < r$ and $x_c + \sqrt{r^2 - y_c^2} > r$.
We aim to demonstrate that the inequality
\begin{equation}
y_c + \sqrt{r^2 - y_c^2} < x_c + \sqrt{r^2 - x_c^2}
\end{equation}
holds if and only if $x_c > y_c$.

We start with the inequality:
\begin{equation}
y_c + \sqrt{r^2 - y_c^2} < x_c + \sqrt{r^2 - x_c^2}.
\end{equation}

Subtracting $y_c$ from both sides, we obtain:
\begin{equation}
\sqrt{r^2 - y_c^2} < x_c - y_c + \sqrt{r^2 - x_c^2}.
\end{equation}
We can further simplify this to:
\begin{equation}
\sqrt{r^2 - y_c^2} - \sqrt{r^2 - x_c^2} < x_c - y_c.
\end{equation}

The inequality now compares two quantities: $\sqrt{r^2 - y_c^2} - \sqrt{r^2 - x_c^2}$ and $x_c - y_c$.
\begin{itemize}
\item The term $\sqrt{r^2 - y_c^2}$ represents the horizontal distance from the point $(x_c, y_c)$ to the vertical axis.
\item The term $\sqrt{r^2 - x_c^2}$ represents the vertical distance from the point $(x_c, y_c)$ to the horizontal axis.
\end{itemize}

Let's consider the case where $x_c > y_c$:
\begin{itemize}
  \item If $x_c > y_c$, then $x_c - y_c > 0$.
  \item Additionally, $\sqrt{r^2 - y_c^2} > \sqrt{r^2 - x_c^2}$ because $y_c < x_c$.
\end{itemize}

This implies that the term $\sqrt{r^2 - y_c^2} - \sqrt{r^2 - x_c^2}$ is positive, and it is less than $x_c - y_c$, proving that the inequality holds under this condition.

Thus, for the inequality $y_c + \sqrt{r^2 - y_c^2} < x_c + \sqrt{r^2 - x_c^2}$ to hold, it is necessary that $x_c > y_c$.
So, the $x$ axis is the closest axis to the center of the sphere and is intersected once.

%%%%%%%%%%%%%%%%%%%%%%%%%%%%%%%%%%%%%%%%%%%%
\paragraph{3D Case}
To convert this proof from 2D to 3D, we need to extend the concepts from the circle and cross to a sphere and a three dimensional cross~\footnote{That is, three axis aligned orthogonal segments of the same size intersecting at their respective centers.}.
Consider a sphere with radius $ r $ centered at $ (x_c, y_c, z_c) $ in 3D space, and a cross (or coordinate axes) centered at the origin with each axis extending from $-l$ to $ l $. 
We are interested in analyzing the intersection of the sphere with the axes, focusing particularly on the first octant where $ 0 \leq x_c \leq l $, $ 0 \leq y_c \leq l $, and $ 0 \leq z_c \leq l $.

The equation of the sphere is given by:
\begin{equation}
(x - x_c)^2 + (y - y_c)^2 + (z - z_c)^2 = r^2.
\end{equation}

The coordinates of the intersection points of the sphere with the axes are found by setting two of the coordinates to zero in the sphere's equation:
\begin{itemize}
  \item Intersection with the $ x $-axis (set $ y = 0 $ and $ z = 0 $):
   \begin{equation}
   x = x_c \pm \sqrt{r^2 - y_c^2 - z_c^2}
   \end{equation}
   
  \item Intersection with the $ y $-axis (set $ x = 0 $ and $ z = 0 $):
   \begin{equation}
   y = y_c \pm \sqrt{r^2 - x_c^2 - z_c^2}
   \end{equation}
   
  \item Intersection with the $ z $-axis (set $ x = 0 $ and $ y = 0 $):
   \begin{equation}
   z = z_c \pm \sqrt{r^2 - x_c^2 - y_c^2}
   \end{equation}
\end{itemize}

Let's denote the intersection points on the positive half of the axes as:
\begin{itemize}
\item $ x_1 = x_c + \sqrt{r^2 - y_c^2 - z_c^2} $
\item $ y_1 = y_c + \sqrt{r^2 - x_c^2 - z_c^2} $
\item $ z_1 = z_c + \sqrt{r^2 - x_c^2 - y_c^2} $
\end{itemize}

We need to analyze whether these points lie within the bounds of the cross, i.e., whether $ x_1 \leq l $, $ y_1 \leq l $, and $ z_1 \leq l $.

Assume that one of these coordinates, say $ x_1 $, exceeds $ l $. This would mean that the intersection does not lie on the cross, i.e., $ x_c + \sqrt{r^2 - y_c^2 - z_c^2} > l $. 
Similarly, for $ y_1 $ and $ z_1 $, we require that:
\begin{equation}
y_c + \sqrt{r^2 - x_c^2 - z_c^2} > l \quad \text{or} \quad z_c + \sqrt{r^2 - x_c^2 - y_c^2} > l.
\end{equation}
These conditions cannot all be true simultaneously for $ x_c $, $ y_c $, and $ z_c $ within the defined range, similar to the 2D case. 
Thus, a sphere will intersect the cross at most once per axis.

Next, we determine the axis closest to the sphere's center. If $ x_c > y_c > z_c $, we aim to prove that the intersection on the $ x $-axis occurs first (i.e., is the smallest).

Starting with:
\begin{equation}
  \begin{split}
& x_c + \sqrt{r^2 - y_c^2 - z_c^2} < y_c + \sqrt{r^2 - x_c^2 - z_c^2} \quad \\ & \text{and} \quad x_c + \sqrt{r^2 - y_c^2 - z_c^2} < z_c + \sqrt{r^2 - x_c^2 - y_c^2}.
\end{split}
\end{equation}
These can be simplified, following similar steps as in the 2D case:
\begin{equation}
  \begin{split}
& \sqrt{r^2 - y_c^2 - z_c^2} - \sqrt{r^2 - x_c^2 - z_c^2} < y_c - x_c \quad \\ & \text{and} \quad \sqrt{r^2 - y_c^2 - z_c^2} - \sqrt{r^2 - x_c^2 - y_c^2} < z_c - x_c.
  \end{split}
\end{equation}

The argument follows that since $ x_c > y_c > z_c $, the inequalities hold true, confirming that the $ x $-axis is the closest, and thus it is intersected first.

The 3D extension of the proof shows that a sphere intersects each axis of a coordinate cross at most once, and the axis closest to the sphere's center (in the order of $ x_c > y_c > z_c $) will have the intersection point closest to the origin.

%%%%%%%%%%%%%%%%%%%%%%%%%%%%%%%%%%%%%%%%%%%%

\section{Conversion from Particles' Positions to Triangles}
\label{app:postotriangles}

\begin{lstlisting}[language=C++, caption={Triangles generation from particles's positions.}, label={lst:parttopos}]
  for(int i = 0; i < nbPoints; i++)
  {
      const float3 point = points[i];
      std::array<float3, 8> corners;
      for(int idxCorner = 0 ; idxCorner < 8 ; ++idxCorner){
          corners[idxCorner].z = point.z + (idxCorner&1 ? (cutoffRadius/2)+epsilon : (-cutoffRadius/2)-epsilon );
          corners[idxCorner].y = point.y + (idxCorner&2 ? (cutoffRadius/2)+epsilon : (-cutoffRadius/2)-epsilon );
          corners[idxCorner].x = point.x + (idxCorner&4 ? cutoffRadius/2 : -cutoffRadius/2 );
      }
      vertices.push_back(corners[0]);
      vertices.push_back(corners[1]);
      vertices.push_back(corners[3]);

      vertices.push_back(corners[0]);
      vertices.push_back(corners[2]);
      vertices.push_back(corners[3]);

      vertices.push_back(corners[4]);
      vertices.push_back(corners[5]);
      vertices.push_back(corners[7]);

      vertices.push_back(corners[4]);
      vertices.push_back(corners[6]);
      vertices.push_back(corners[7]);
  }
  \end{lstlisting}

\section{Generation of particles on a sphere}
\label{app:gensurface}

\begin{lstlisting}[language=C++, caption={Non-uniform test cas generation.}, label={lst:gensurface}]
    const int MaxParticlesPerCell = 32;
    const int MaxBoxDiv = 128;
    for(int boxDiv = 2 ; boxDiv <= MaxBoxDiv ; boxDiv *= 2){
        const int nbBoxes = boxDiv*boxDiv*boxDiv;
        for(int nbParticles = nbBoxes ; nbParticles <= nbBoxes*MaxParticlesPerCell ; nbParticles *= 2){
            const double particlePerCell = double(nbParticles)/double(nbBoxes);
            const double expectedNbNeighbors = 9*particlePerCell;
            const double coef = 1. - ((2*expectedNbNeighbors)/nbParticles);
            const double validCoef = std::min(1.0, std::max(-1.0, coef));
            const double sphereRadius = acos(validCoef);

            // The following is only used if we need to build a grid of cell
            const float boxWidth = std::ceil(2.0 / sphereRadius) * sphereRadius;
            const int gridDim = boxWidth/sphereRadius;
            const float cellWidth = boxWidth/gridDim;
    ......

    auto generateRandomParticle() {
        double theta = 2.0 * M_PI * ((double)rand() / RAND_MAX); // Random angle between 0 and 2PI
        double phi = acos(1.0 - 2.0 * ((double)rand() / RAND_MAX)); // Random angle between 0 and PI
    
        // Convert spherical coordinates to Cartesian coordinates
        double x = sin(phi) * cos(theta);
        double y = sin(phi) * sin(theta);
        double z = cos(phi);
    
        return Particle{x, y, z};
    }
  \end{lstlisting}

\end{document}